\documentclass[a4paper,11pt]{article}
\usepackage[dvipdfmx]{graphicx}
\usepackage{amsfonts}
\usepackage{physics,amsmath}
\usepackage{amssymb}
\usepackage{mathtools}
\usepackage{colortbl}
\usepackage{cite}
\usepackage{here}
\usepackage{hyperref}
\usepackage{mathrsfs}
\usepackage{algpseudocode}
\usepackage{xcolor}
\usepackage{ulem}
\usepackage{comment}
\numberwithin{equation}{section}
\usepackage[left=2cm,right=2cm,top=3.5cm,bottom=3.5cm]{geometry}

\allowdisplaybreaks[1]
\hypersetup{
	colorlinks=true,
	linkcolor=ceruleanblue,
	filecolor=ceruleanblue,      
	urlcolor=ceruleanblue,
	citecolor=ceruleanblue,
}
\definecolor{ceruleanblue}{rgb}{0.0, 0.2, 0.6}
\newcommand{\de}{{\rm d}}

\date{\today}

\begin{document}

\begin{flushright} {\footnotesize YITP-24-62, IPMU24-0018, RIKEN-iTHEMS-Report-24}  \end{flushright}

\begin{center}
\LARGE{\bf Tidal Love Numbers from EFT of Black Hole Perturbations with Timelike Scalar Profile}
\\[1cm] 

\large{Chams Gharib Ali Barura$^{\,\rm a}$, Hajime Kobayashi$^{\,\rm b}$, Shinji Mukohyama$^{\,\rm b, \rm c}$, Naritaka Oshita$^{\,\rm b, \rm d, \rm e}$, Kazufumi Takahashi$^{\,\rm b}$, and Vicharit Yingcharoenrat$^{\,\rm c, \rm f}$}
\\[0.5cm]

\small{
\textit{$^{\rm a}$ The \'{E}cole Normale Sup\'{e}rieure Paris-Saclay, 4 Av.~des Sciences, 91190 Gif-sur-Yvette, France}}
\vspace{.2cm}

\small{\textit{$^{\rm b}$
Center for Gravitational Physics and Quantum Information, Yukawa Institute for Theoretical Physics, 
\\ Kyoto University, 606-8502, Kyoto, Japan}}
\vspace{.2cm}

\small{
\textit{$^{\rm c}$
Kavli Institute for the Physics and Mathematics of the Universe (WPI), The University of Tokyo Institutes for Advanced Study (UTIAS), The University of Tokyo, Kashiwa, Chiba 277-8583, Japan}}
\vspace{.2cm}

\small{
\textit{$^{\rm d}$
The Hakubi Center for Advanced Research, Kyoto University, Yoshida Ushinomiyacho, Sakyo-ku, Kyoto 606-8501, Japan}}
\vspace{.2cm}

\small{
\textit{$^{\rm e}$
RIKEN iTHEMS, Wako, Saitama, 351-0198, Japan}}
\vspace{.2cm}

\small{
\textit{$^{\rm f}$
High Energy Physics Research Unit, Department of Physics, Faculty of Science, Chulalongkorn University, 254 Phayathai Road, Pathumwan, Bangkok 10330, Thailand}}
\vspace{.2cm}
\end{center}

\vspace{0.3cm} 

\begin{abstract}\normalsize
We study static tidal Love numbers (TLNs) of a static and spherically symmetric black hole for odd-parity metric perturbations.
We describe black hole perturbations using the effective field theory (EFT), formulated on an arbitrary background with a timelike scalar profile in the context of scalar-tensor theories.
In particular, we obtain a static solution for the generalized Regge-Wheeler equation order by order in a modified-gravity parameter and extract the TLNs uniquely by analytic continuation of the multipole index $\ell$ to non-integer values.
For a stealth Schwarzschild black hole, the TLNs are vanishing as in the case of Schwarzschild solution in general relativity.
We also study the case of Hayward black hole as an example of non-stealth background, where we find that the TLNs are non-zero (or there is a logarithmic running).
This result suggests that our EFT allows for non-vanishing TLNs and can in principle leave a detectable imprint on gravitational waves from inspiralling binary systems, which opens a new window for testing gravity in the strong-field regime.
\end{abstract}

\vspace{0.3cm} 

\vspace{2cm}

\newpage
{
\hypersetup{linkcolor=black}
\tableofcontents
}

\flushbottom

\vspace{1cm}


\section{Introduction}\label{sec:Introduction}
With the direct detection of gravitational waves (GWs) by LIGO-Virgo-KAGRA collaborations~\cite{LIGOScientific:2016aoc,LIGOScientific:2018mvr,LIGOScientific:2020ibl,LIGOScientific:2021usb,KAGRA:2021vkt}, there are increasing opportunities to obtain observational information about the nature of gravity in the strong-field regime. 
GW signals from the inspiral phase of a compact binary merger, such as black holes (BHs) or neutron stars, are one of the most promising observational targets. 
During the inspiral phase when the orbital separation of the binary is large, the binary components are well approximated by point particles~\cite{Blanchet:2013haa}.
The back-reaction of the GW emission causes the orbital separation to shrink, and towards the end of the inspiral phase, finite-size effects become relevant.
Among them, the response of an object to external force is crucial, because it reflects the internal structure of the object as well as the underlying theory of both gravity and matter.
In particular, the deformability of a compact object is quantified by the so-called {\it tidal Love numbers} (TLNs)~\cite{Hinderer:2007mb, Damour:2009vw,Binnington:2009bb}, 
and the effects of tidal deformation are imprinted in the gravitational waveform at 5 post-Newtonian order~\cite{Flanagan:2007ix,Damour:2012yf,Favata:2013rwa}.
Therefore, it is possible to test the structure of dense matter and strong-field gravity by measuring the TLNs from the observed signals of GWs~\cite{Cardoso:2017cfl,Katagiri:2023umb}.
Using the information on the TLNs, the event~GW170817 can constrain the equation of state of the neutron star interior~\cite{LIGOScientific:2017apx,LIGOScientific:2018cki,LIGOScientific:2018hze}.
In general relativity (GR), the unique stationary, axisymmetric, and asymptotically flat solution of the vacuum Einstein equation with a regular horizon is given by the Kerr metric, for which the TLNs vanish in four spacetime dimensions~\cite{Damour:2009vw,Binnington:2009bb,Kol:2011vg,Hui:2020xxx,LeTiec:2020bos,LeTiec:2020spy,Chia:2020yla,Charalambous:2021mea}.
On the other hand, BHs in theories beyond four-dimensional GR can have non-vanishing TLNs~\cite{Cardoso:2017cfl,Cardoso:2018ptl,Chakravarti:2018vlt,Cardoso:2019vof,Brown:2022kbw,DeLuca:2022tkm}.
Also, horizonless compact objects that may arise due to quantum gravity correction~\cite{Uchikata:2016qku,Cardoso:2017cfl,Addazi:2018uhd,Maselli:2018fay,Cardoso:2019rvt,Cardoso:2019nis,Chakraborty:2023zed} and BHs surrounded by matter fields have non-zero TLNs in general~\cite{Cardoso:2019upw,DeLuca:2021ite,DeLuca:2022xlz,Katagiri:2023yzm,Katagiri:2023umb}.
In addition, it is known that there are cases where TLNs can have a logarithmic scale dependence, which indicates the ``running'' of TLNs (see, e.g.,~\cite{Cardoso:2019upw,Hui:2020xxx,DeLuca:2022tkm,Katagiri:2023umb} and references therein).
From a theoretical point of view, it is important to clarify the origin of the (non-)vanishing or the running of TLNs.\footnote{From the viewpoint of the point-particle (worldline) EFT~\cite{Kol:2011vg,Ivanov:2022hlo}, the logarithmic scale dependence of TLNs is understood as classical renormalization group running. It is yet unclear under which conditions the logarithmic running is absent. (This is a kind of {\it naturalness problem} as pointed out in \cite{Porto:2016zng,Charalambous:2021mea,Charalambous:2022rre}.)}
It was recently suggested in \cite{Hui:2021vcv,BenAchour:2022uqo,Hui:2022vbh,Charalambous:2022rre,Katagiri:2022vyz,Berens:2022ebl,Sharma:2024hlz} that the vanishing of TLNs in GR is due to the hidden {\it ladder} symmetries.
Therefore, a discovery of non-zero TLNs for BHs can be a smoking gun for new physics beyond GR at the horizon scale.

When one considers theories of gravity beyond GR, one typically introduces new field(s) other than the metric tensor, at least effectively.
Among them, scalar-tensor theories, which include just one scalar field in addition to the metric, have long been studied in the context of both cosmology and compact objects such as BHs and neutron stars.
The most general class of scalar-tensor theories in four-dimensional spacetime with second-order covariant Euler-Lagrange equations is known as Horndeski theory~\cite{Horndeski:1974wa,Deffayet:2011gz,Kobayashi:2011nu}, unifying those traditional scalar-tensor theories, e.g., Brans-Dicke~\cite{Brans:1961sx} and k-essence theories~\cite{ArmendarizPicon:2000dh,ArmendarizPicon:2000ah}.
The second-order nature of Euler-Lagrange equations ensures the absence of Ostrogradsky ghosts~\cite{Woodard:2015zca}, but still one can go beyond the Horndeski class, keeping the ghost-freeness:
Even if the Euler-Lagrange equations involve higher-order derivatives, one can circumvent the problem of Ostrogradsky ghosts by imposing the so-called degeneracy conditions~\cite{Motohashi:2014opa,Langlois:2015cwa,Motohashi:2016ftl,Klein:2016aiq,Motohashi:2017eya,Motohashi:2018pxg}, and scalar-tensor theories respecting such conditions are called degenerate higher-order scalar-tensor (DHOST) theories~\cite{Langlois:2015cwa,Crisostomi:2016czh,BenAchour:2016fzp,Takahashi:2017pje,Langlois:2018jdg}.
The DHOST theory was then extended to the U-DHOST theory~\cite{DeFelice:2018ewo,DeFelice:2021hps, DeFelice:2022xvq}, which involves an instantaneous mode (also known as the shadowy mode) due to the fact that the degeneracy conditions are satisfied only in the unitary gauge.
A yet further extension can be obtained by applying a higher-derivative generalization of the invertible disformal transformations~\cite{Takahashi:2021ttd,Takahashi:2023vva} to Horndeski theories and U-DHOST theories, which is known as the generalized disformal Horndeski~\cite{Takahashi:2022mew} and generalized disformal unitary-degenerate theories~\cite{Takahashi:2023jro}.\footnote{Note that the number of degrees of freedom is not changed by an invertible transformation~\cite{Domenech:2015tca,Takahashi:2017zgr}. 
If two gravitational theories are related to each other via invertible transformation, they can be distinguished only when the matter sector is taken into account.
This is because matter fields define a special frame where the coupling to gravity is minimal (i.e., the Jordan frame). The class of generalized disformal theories allowing for consistent matter coupling was discussed in \cite{Takahashi:2022mew,Naruko:2022vuh,Takahashi:2022ctx,Ikeda:2023ntu}.}

Given that there are so many scalar-tensor theories proposed so far, it would be nice to have a framework that allows us to make phenomenological predictions in a model-independent manner, and the effective field theory (EFT) approach is ideal for this purpose.
The field content, the symmetry breaking pattern, and the background spacetime are the three main components of the EFT. 
In \cite{Arkani-Hamed:2003pdi,Arkani-Hamed:2003juy}, an EFT of scalar-tensor theories on Minkowski and de Sitter spacetimes, called ghost condensation, was formulated. It assumes that the time diffeomorphism invariance is spontaneously broken by a timelike gradient of the scalar field and that the EFT remains invariant under the spatial diffeomorphism, the time translation, and the time reflection, up to the shift and the reflection of the scalar field.
This EFT approach was subsequently extended to a Friedmann-Lema{\^i}tre-Robertson-Walker background, which is known as the EFT of inflation/dark energy~\cite{Cheung:2007st,Gubitosi:2012hu}.
The EFT with a timelike scalar profile has recently been generalized to an arbitrary background geometry~\cite{Mukohyama:2022enj}.
A dictionary that provides a connection between the EFT and concrete covariant theories was also constructed, which applies to Horndeski~\cite{Mukohyama:2022enj} and shift-symmetric quadratic higher-order scalar-tensor (HOST) theories~\cite{Mukohyama:2022skk}. 
The EFT of \cite{Mukohyama:2022enj} can in principle describe physics at both cosmological and BH scales in a single framework because of the timelike nature of the scalar field, which offers the possibility of extracting some information about scalar-field dark energy from GW observations of astrophysical BHs.\footnote{Note that, in \cite{Franciolini:2018uyq}, an EFT of BH perturbations with a spacelike scalar profile on a static and spherically symmetric background was constructed (see also \cite{Hui:2021cpm} for the formulation of the EFT for a slowly rotating BH).}
Moreover, as a first phenomenological application of this EFT, in \cite{Mukohyama:2023xyf,Konoplya:2023ppx}, the spectrum of quasinormal modes was studied in detail.

Among many applications of the EFT, our main goal in this paper is to extract the static TLNs on a static and spherically symmetric BH background.
Here, for simplicity, we do not include the effects of matter fields.\footnote{
In an actual case, the deformation of the black hole may backreact on the matter sector leading to the external field, but we assume that such an effect is tiny and can be ignored.
}
In this setup, it is useful to decompose perturbations into odd- and even-parity modes as they evolve independently in models without parity violation.
In \cite{Mukohyama:2022skk}, the master equation, dubbed the generalized Regge-Wheeler (RW) equation, for odd-parity perturbations was derived using the EFT.
In this paper, we will use it to compute the static TLNs of the odd-parity perturbations.

Having said that, it is non-trivial how to define TLNs of BHs.
In general, a static solution for perturbations that is regular at the BH horizon is a linear combination of growing and decaying modes at spatial infinity, and a set of TLNs is usually defined as the ratio between the amplitudes of these two modes.
The problem is that the decaying mode can be degenerate with subleading terms of the growing mode in the far region, which introduces an ambiguity in the definition of TLNs.\footnote{In the Newtonian limit, there is no such ambiguity to define the TLNs, because solutions of the Laplace equation have no subleading terms~\cite{Poisson2014gravity}.}
As discussed in \cite{Gralla:2017djj}, resolving this ambiguity would be important for an accurate prediction of gravitational waveforms.
A possible solution for this ambiguity is to perform an analytic continuation of the multipole index~$\ell$ from integer values to generic complex values~\cite{Kol:2011vg,LeTiec:2020bos,Creci:2021rkz}.
In this paper, we formulate a general method to find a static solution of the generalized RW equation with a regular boundary condition at the horizon and to extract the TLNs unambiguously by promoting the parameter~$\ell$ to non-integer values in the intermediate steps of our calculations.
For illustrative purposes, we consider two examples of hairy BHs: One is the stealth Schwarzschild BH and the other is the Hayward BH with a non-trivial scalar field. 
In the former, the TLNs are vanishing as in GR. On the other hand, for the latter which is an example of non-stealth solutions, we show that the TLNs are non-vanishing and that a logarithmic running of TLNs shows up at some value of $\ell$.

The structure of this paper is as follows. In Section~\ref{sec:setup}, we revisit the EFT of perturbations on an arbitrary background with a timelike scalar profile and provide a brief review of the master equation (i.e., the generalized RW equation) for odd-parity perturbations on a static and spherically symmetric BH.
Section~\ref{sec:analysis} is the main part of this paper. First, we review the relativistic definition of the TLNs for (stealth) Schwarzschild BHs.
Then, we develop a general formalism to calculate the TLNs based on the generalized RW equation for generic non-Schwarzschild backgrounds and demonstrate it with the Hayward background.
Finally, in Section~\ref{sec:discussion}, we draw our conclusions and prospects of our formalism.

\section{Setup}\label{sec:setup}

\subsection{Formulation of the EFT: a review}

Here, we briefly review the formulation of the EFT for perturbations on an arbitrary background with a timelike scalar profile~\cite{Mukohyama:2022enj,Mukohyama:2022skk}.
The key idea of our EFT is that the scalar field~$\Phi$ has a timelike background profile, $\bar{\Phi}$, which spontaneously breaks the time diffeomorphism invariance and defines a preferred ($\Phi = {\rm const.}$) slicing whose unit normal vector can be defined by
\begin{align}\label{eq:normal_EFT}
n_\mu = 
-\frac{\partial_\mu \Phi}{\sqrt{-X}}
\rightarrow - \frac{\delta_\mu^\tau}{\sqrt{-g^{\tau\tau}}} \;,
\end{align}
with $X \equiv g^{\mu\nu}\partial_\mu\Phi\partial_\nu\Phi\,(<0)$, satisfying $n_\mu n^\mu = -1$. 
Here, we use $\tau$ to denote the time coordinate where $\bar{\Phi}=\bar{\Phi}(\tau)$, and the arrow in Eq.~\eqref{eq:normal_EFT} refers to taking the unitary gauge, where $\delta\Phi = \Phi-\bar{\Phi}(\tau) = 0$.
The spontaneous breaking of time diffeomorphism invariance by $\bar{\Phi}$ implies that the unitary-gauge EFT action remains invariant under the 3d diffeomorphism but that it does not respect the time diffeomorphism invariance.
Therefore, the EFT action may contain a scalar function of the 4d and 3d curvatures, the extrinsic curvature, the ($\tau\tau$)-component of the inverse metric tensor, the time coordinate~$\tau$, and so on. 

For convenience, we write down the metric using the Arnowitt-Deser-Misner (ADM) $3+1$ decomposition as
\begin{align}
\de s^2=-N^2\de\tau^2+h_{ij}(\de x^i+N^i\de\tau)(\de x^j+N^j\de\tau)\;,
\end{align}
where $N$ denotes the lapse function, $N^i$ the shift vector, and $h_{ij}$ the induced metric on each constant-$\tau$ hypersurface.
The extrinsic curvature of the hypersurface and its trace are given by\footnote{Note that the extrinsic curvature can be determined by the projection tensor~$h_{\mu\nu} = g_{\mu\nu} + n_\mu n_\mu$ and the normal vector~$n^\mu$: $K_{\mu\nu} = h_\mu^\rho\nabla_\rho n_\nu$ with $\nabla_\mu$ being the 4d covariant derivative.}
\begin{align}
K_{ij} = \frac{1}{2N}(\Dot{h}_{ij}-{\rm D}_iN_j-{\rm D}_jN_i) \;, \qquad K=h^{ij}K_{ij}\;,
\end{align}
where a dot denotes the derivative with respect to $\tau$, and ${\rm D}_i$ denotes the covariant derivative associated with $h_{ij}$. 
As usual, it is straightforward to compute other geometrical quantities such as the 3d Ricci tensor~${}^{(3)}\!R_{ij}$.
Note that the 3d Weyl tensor identically vanishes, and hence the 3d Riemann tensor can be written in terms of the 3d Ricci tensor.

Let us now briefly review the construction of the EFT, following \cite{Mukohyama:2022enj}. 
As explained before, in the unitary gauge, the EFT action can depend on a scalar function, made out of the 3d quantities such as $K_{ij}$ and ${}^{(3)}\!R$, in addition to the 4d covariant objects. Also, it can contain an explicit dependence on $\tau$. Therefore, the general unitary-gauge EFT action has the following form:
\begin{align}
S=\int \de^4x \sqrt{-g}\,L({}^{(3)}\!R^\mu_\nu, K^\mu_\nu, g^{\tau\tau}, \tau,{\nabla}_\mu)\;, \label{S_EFT}
\end{align}
where $L$ is a scalar function. Note that $g^{\tau\tau}$ is treated as an independent EFT building block. We emphasize that this EFT action can be applied to an arbitrary background geometry.
In order to obtain the EFT of perturbations, one then expands the action~(\ref{S_EFT}) around a non-trivial background. 
For instance, we define $\delta g^{\tau\tau} \equiv g^{\tau\tau} - \bar{g}^{\tau\tau}(\tau, \vec{x})$ and $\delta K^\mu_\nu \equiv K^\mu_\nu - \bar{K}^\mu_\nu(\tau, \vec{x})$, where the background values~$\bar{g}^{\tau\tau}$ and $\bar{K}^\mu_\nu$ can depend on both $\tau$ and $\vec{x}$, as opposed to the case of EFT of inflation/dark energy. Then, expanding the action~(\ref{S_EFT}) order by order in perturbations, we obtain the EFT action that can be practically used to describe dynamics of perturbations.
At each order in perturbations, the EFT is characterized by a finite number of functions of $(\tau,\vec{x})$ multiplying by the perturbations, which we refer to as EFT parameters.

It should be noted that one cannot freely choose the EFT parameters, to be consistent with the 3d diffeomorphism invariance of our EFT.
As explained in detail in \cite{Mukohyama:2022enj,Mukohyama:2022skk}, each term of the Taylor expansion of the action~(\ref{S_EFT}) breaks the 3d diffeomorphism invariance due to the inhomogeneities of the background quantities, e.g., $\bar{K}^\mu_\nu(\tau, \vec{x})$. However, the fact that the action~(\ref{S_EFT}) is manifestly invariant under the spatial diffeomorphism implies that there must be relations among the Taylor coefficients that ensure the 3d diffeomorphism invariance order by order in perturbations.
These relations are called consistency relations, associated with the spatial diffeomorphism invariance.
Without entering into further detail, one can straightforwardly obtain the consistency relations by applying the chain rule with respect to the $\vec{x}$-derivatives.  
This results in a set of infinitely many consistency relations among the EFT coefficients.
Imposing these relations on the EFT parameters, we obtain the consistent EFT action in the unitary gauge.
We note that the relations obtained by applying the chain rule with respect to the $\tau$-derivatives are automatically satisfied due to the explicit $\tau$ dependence of the EFT Lagrangian.
If one imposes the shift symmetry to the EFT action, the explicit $\tau$ dependence is prohibited, and hence one obtains an additional set of consistency relations~\cite{Finelli:2018upr,Khoury:2022zor}.\footnote{In the context of EFT of vector-tensor theories~\cite{Aoki:2021wew,Aoki:2023bmz} where the shift symmetry is gauged, one obtains a set of consistency relations corresponding to the combined $U(1)$ and the time diffeomorphism.}
Finally, the 4d diffeomorphism covariance of the EFT can be recovered by the usual St\"{u}ckelberg trick: $\tau \rightarrow \tau + \pi(\tau,\vec{x})$, with $\pi$ being a Nambu-Goldstone boson associated with the broken time diffeomorphism invariance.
In the next Sections, we will describe the background dynamics and the odd-parity perturbations based on the EFT.

\subsection{Static and spherically symmetric background}

Let us consider a static, spherically symmetric, asymptotically flat metric as the background spacetime:
\begin{align}
\de s^2 &= \bar{g}_{\mu\nu}\de x^\mu \de x^\nu 
= -A(r)\de t^2 + \frac{\de r^2}{B(r)} + r^2\de\Omega_2^2\;,
\end{align}
where $\de\Omega_2^2 = \de\theta^2 + \sin^2\theta \de\varphi^2$ represents the line element on a unit 2-sphere.
Such a static background metric can be compatible with a linearly time-dependent scalar field, $\bar{\Phi} = \mu^2 t + \phi(r)$ with $\mu^2$ being a non-vanishing constant, if the action has a shift symmetry with respect to the scalar field.
Assuming that the scalar field has a timelike gradient, i.e., $\bar{X} = \bar{g}^{\mu\nu}\partial_\mu\bar{\Phi}\partial_\nu\bar{\Phi}<0$, our EFT applies.
In the present paper, for simplicity, we only consider the case where $\bar{X}={\rm const}$.
Actually, most of the stealth solutions known so far has a constant $\bar{X}$.
In this case, the vector~$u^\mu = \bar{g}^{\mu\nu}\partial_\nu\bar{\Phi}$ is tangent to the geodesic, as $u^\nu\nabla_\nu u^\mu \propto \nabla^\mu\bar{X} = 0$. Moreover, since $u^\mu$ is timelike, $\bar{\Phi}$ corresponds to the proper time for a congruence of freely falling observers.
Thus, using the coordinate system synchronized with a congruence of freely falling observers (referred to as Lema{\^i}tre coordinates~\cite{Lemaitre:1933gd,Mukohyama:2005rw,Khoury:2020aya,Takahashi:2021bml}), given by
\begin{align}
\de s^2 &= -\de\tau^2 + [1 - A(r)]\de\rho^2 + r^2\de\Omega_2^2\;, \label{bkg-metric}
\end{align}
we can express $\bar{\Phi} \propto \tau$ and $\bar{g}^{\tau\tau} = -1$.
The relation between $(t,r)$ and $(\tau,\rho)$ coordinates is given by 
\begin{align}
\de\tau = \de t + \sqrt{\frac{1 - A}{AB}}\,\de r\;, \qquad
\de\rho = \de t + \frac{\de r}{\sqrt{AB(1 - A)}}\;.
\end{align}
The areal radius~$r$ is now a function of $\rho - \tau$, and we have the relation~$\partial_\rho r = -\dot{r} = \sqrt{B(1 - A)/A}$.

Let us proceed to the background equations of motion. Assuming that we can move to the frame where the coefficient in front of the Ricci scalar is a constant, the general EFT action relevant to background dynamics is given by\footnote{We note that the coefficient in front of $K^\nu_\mu$ is assumed to be proportional to $\bar{K}^\mu_\nu$ for simplicity, as it is the case for shift- and reflection-symmetric HOST theories.}
\begin{align}
		S_{\rm tadpole} = \int \de^4x \sqrt{-g} \bigg[&\frac{M_\star^2}{2} R - \Lambda(r) - c(r)g^{\tau\tau} - \tilde{\beta}(r) K - \alpha(r)\bar{K}^\mu_\nu K^{\nu}_\mu -\zeta(r) n^\mu\partial_\mu g^{\tau\tau} 
		\bigg] \;,
	\label{eq:EFT_HOST}
\end{align}
where we have defined 
\begin{align}
R \equiv {}^{(3)}\!R + K_{\mu\nu} K^{\mu\nu} - K^2 = \tilde{R} - 2 \nabla_\mu (K n^\mu - n^\nu \nabla_\nu n^\mu) \;, 
\end{align}
with $\tilde{R}$ being the standard 4d Ricci scalar.
Here, for simplicity, terms involving $r^{\mu}_{\nu}\equiv {}^{(3)}\!R^{\mu}_{\nu}-{}^{(3)}\!Rh^{\mu}_{\nu}/3$ have been omitted. 
Notice that the EFT coefficients depend only on the radial coordinate~$r$ due to the shift symmetry of the scalar field and the static and spherically symmetric nature of the background fields.
The background equations of motion (or the tadpole cancellation conditions) are straightforwardly obtained by varying the action~\eqref{eq:EFT_HOST} with respect to the metric, which can be written in the form
\begin{align}
M_\star^2\bar{G}_{\mu\nu}&=\bar{T}_{\mu\nu}\;.\label{Tadpole}
\end{align}
Here, $\bar{G}_{\mu\nu}$ is the Einstein tensor calculated from the background metric $\bar{g}_{\mu\nu}$, and $\bar{T}_{\mu\nu}$ is the energy-momentum tensor given by
\begin{align}
	\bar{T}_{\mu\nu}&=
	-({\Lambda}-c-\bar{n}^{\lambda}\partial_{\lambda}\tilde{\beta}+\alpha\bar{K}^{\lambda}_\sigma\bar{K}^\sigma_{\lambda})\bar{g}_{\mu\nu}+[2c-\bar{n}^{\lambda}\partial_{\lambda}\tilde{\beta}+\alpha\bar{K}^{\lambda}_\sigma\bar{K}^\sigma_{\lambda}-2\bar{\nabla}_{\lambda}(\zeta\bar{n}^{\lambda})]\bar{n}_\mu\bar{n}_\nu \nonumber \\
	&\quad -2\bar{n}_{(\mu}\partial_{\nu)}\tilde{\beta}+2\alpha\bar{K}_\mu^{\lambda}\bar{K}_{\lambda\nu}-2\bar{\nabla}_{\lambda}(\alpha\bar{K}^{\lambda}_{(\mu}\bar{n}_{\nu)})+\bar{\nabla}_{\lambda}(\alpha\bar{K}_{\mu\nu}\bar{n}^{\lambda})\;. \label{eq:T_munu}
\end{align}
The non-vanishing components of the Einstein tensor are given by
\begin{equation}
\begin{split}
    \bar{G}^\tau{}_\tau&=-\frac{[r(1-B)]'}{r^2}+\frac{1-A}{r}\left(\frac{B}{A}\right)'\;,\qquad
    \bar{G}^\tau{}_\rho=-\frac{1-A}{r}\left(\frac{B}{A}\right)'\;,\\
    \bar{G}^\rho{}_\rho&=-\frac{[r(1-B)]'}{r^2}-\frac{1}{r}\left(\frac{B}{A}\right)'\;,\qquad 
    \bar{G}^\theta{}_\theta=\frac{B(r^2A')'}{2r^2A}+\frac{(r^2A)'}{4r^2}\left(\frac{B}{A}\right)'\;,
\end{split}
\end{equation}
where a prime denotes differentiation with respect to $r$.
Therefore, using (\ref{eq:T_munu}) in (\ref{Tadpole}), we obtain~\cite{Mukohyama:2022skk}
\begin{equation}
\begin{split}
    &\Lambda-c	=M_\star^2(\bar{G}^\tau{}_\rho-\bar{G}^\rho{}_\rho)\;, \\
    &\Lambda+c+\frac{2}{r^2}\sqrt{\frac{B}{A}}\left(r^2\sqrt{1-A}\,\zeta\right)'
    =-M_\star^2\bar{G}^\tau{}_\tau\;, \\
    &\left[\partial_\rho\bar{K}+\frac{1-A}{r}\left(\frac{B}{A}\right)'\,\right]\alpha+\frac{A'B}{2A}\alpha'+\sqrt{\frac{B(1 - A)}{A}}\tilde{\beta}'
    =-M_\star^2\bar{G}^\tau{}_\rho\;, \\
    &\frac{1}{2r^2}\sqrt{\frac{B}{A}}\left[r^4\sqrt{\frac{B}{A}}\left(\frac{1-A}{r^2}\right)'\alpha\right]'
    =M_\star^2(\bar{G}^\rho{}_\rho-\bar{G}^\theta{}_\theta)\;.
\end{split} \label{EOM_BG}
\end{equation}
Given the metric functions~$A(r)$ and $B(r)$, Eq.~(\ref{Tadpole}) provides a set of relations among the EFT parameters. In particular, when $A(r)=B(r)$, from the fourth equation in (\ref{EOM_BG}), it follows that~\cite{Mukohyama:2023xyf}
\begin{align}
\alpha=M_\star^2+\frac{3C}{r(2-2A+rA')}\;,
\label{tadpole-A=B-integtated}
\end{align}
with $C$ being a constant.

\subsection{Odd-parity perturbations}\label{sec:odd-parity}

Here, we analyze the dynamics of odd-parity perturbations with $\ell \geq 2$ at the linear level around the background~(\ref{bkg-metric}). The main result of this Section can be found in \cite{Mukohyama:2022skk}.

Thanks to the $SO(2)$ invariance of the metric~(\ref{bkg-metric}), the perturbations can be decomposed into odd- and even-parity modes, using the basis of spherical harmonics. 
Note that there is no mixing of the two modes in the absence of parity-violating couplings.
In this analysis, we focus on the parity-odd modes.
Denoting the metric perturbation as $\delta g_{\mu\nu} = g_{\mu\nu} - \bar{g}_{\mu\nu}$,
the parity-odd part~$\delta g_{\mu\nu}^\mathrm{odd}$ can be expressed in the $(\tau,\rho,\theta,\varphi)$ coordinates as:
\begin{align}\label{eq:odd_pert}
    &\delta g_{\mu\nu}^\mathrm{odd}
    =\sum_{\ell,m} r^2 \left(\begin{array}{ccc}
         0&0& h_{0} E_a^{\ b} \bar{\nabla}_b \\
         0&0&h_{1} E_a^{\ b} \bar{\nabla}_b \\
         h_{0} E_a^{\ b} \bar{\nabla}_b & h_{1}E_a^{\ b} \bar{\nabla}_b & h_{2}E_{(a|}^{\ \ \ c} \bar{\nabla}_c \bar{\nabla}_{|b)}
    \end{array}\right)Y_{\ell m}(\theta, \varphi)\;, 
\end{align}
where $a,b,c\in\{\theta,\varphi\}$ and $h_i \equiv h_i^{\ell m}(\tau,\rho)$ for $i\in\{0,1,2\}$.
Here, $E_{ab}$ represents the completely antisymmetric tensor on the unit 2-sphere and $\bar{\nabla}_a$ denotes the covariant derivative with respect to the unit 2-sphere metric~$\de\Omega_2^2$.
By introducing an infinitesimal coordinate transformation~$x^\mu\to x^\mu+\xi^\mu$ with $\xi^\tau=\xi^\rho=0$ and $\xi^a=\sum_{\ell,m} \Xi_{\ell m}(\tau,\rho) E^{ab} \bar{\nabla}_b Y_{\ell m}(\theta,\varphi)$, we find that $h_2$ can be eliminated using this gauge freedom.
Note that this is a complete gauge fixing and hence can be imposed at the level of Lagrangian~\cite{Motohashi:2016prk}.

In the odd sector, the relevant EFT action up to the quadratic order is given by
\begin{align}
	S_{2} = \int \de^4x \sqrt{-g} \bigg[\frac{M_\star^2}{2}R - \Lambda(r) - c(r)g^{\tau\tau} -\tilde{\beta}(r) K - \alpha(r)\bar{K}^{\mu}_\nu K^\nu_{\mu} + \frac{1}{2} M_3^2(r) \delta K^\mu_\nu \delta K^\nu_\mu \bigg] \;. \label{eq:EFT_action}
	\end{align}
Since the derivation of the generalized RW equation was presented in detail in \cite{Mukohyama:2022skk} (see also \cite{Mukohyama:2023xyf}), here we only explain the main idea and the relevant quantities, which will be used in the next Section.

Plugging the perturbations~(\ref{eq:odd_pert}) into the action~(\ref{eq:EFT_action}) and using the background equations~\eqref{EOM_BG}, it is straightforward to obtain the quadratic action of $h_0$ and $h_1$.
Then, requiring that the coefficient of the term~$h_1\partial_\rho h_0$ vanishes,\footnote{This requirement can be automatically realized in shift- and $Z_2$-symmetric quadratic HOST theories. Away from this class of theories, such a term can show up in general as it is not prohibited by the symmetry of the EFT. Nevertheless, we impose the requirement since the presence of the term~$h_1\partial_\rho h_0$ leads to the absence of a slowly rotating BH solution or otherwise the divergence of the radial sound speed at $r=\infty$~\cite{Mukohyama:2022skk}.}
one finally arrives at the quadratic action of the following form:
\begin{align}\label{eq:chi_s}
\frac{(\ell-1)(\ell+2)(2\ell + 1)}{2 \pi \ell(\ell+1)} \mathcal{L}_2 = s_1 (\partial_\tau\chi)^2 - s_2 (\partial_\rho\chi)^2 - s_3 \chi^2\;,
\end{align}
where the master variable~$\chi$ is written in terms of (the derivatives of) $h_0$ and $h_1$~\cite{Mukohyama:2022skk}.\footnote{Conversely, as shown in Eq.~(5.10) of \cite{Mukohyama:2022skk}, the metric perturbations~$h_0$ and $h_1$ can be expressed in terms of $\chi$ and its derivatives.}
Here, each coefficient~$s_i$ is
\begin{equation}
	\begin{split}
		s_1 
		= \frac{(M_\star^2 + M_3^2)^{2}r^6}{2\sqrt{1 - A}\,M_\star^2} \;, \qquad s_2 = \frac{(M_\star^2 + M_3^2)r^6}{2 (1 - A)^{3/2}}   \;, \qquad 
		s_3 = \ell(\ell+1) \frac{(M^2_\star + M_3^2) r^4}{2\sqrt{1 - A}} + {\cal O}(\ell^0)\;.
\end{split}\label{coeff-of-L2}
\end{equation}
The absence of ghost/gradient instability at the linear level requires $M_\star^2>0$ and $M_\star^2+M_3^2>0$~\cite{Mukohyama:2022skk}.

The radial and angular propagation speeds of $\chi$ are given by
\begin{align}
\begin{split}
c_\rho^2\equiv  \frac{\bar{g}_{\rho\rho}}{|\bar{g}_{\tau\tau}|}\frac{s_2}{s_1}=\frac{M_\star^2}{M_\star^2+M_3^2}\;, \qquad 
c_\theta^2\equiv \lim_{\ell\to \infty}\frac{r^2}{|\bar{g}_{\tau\tau}|}\frac{s_3}{\ell(\ell+1) s_1}=\frac{M_\star^2}{M_\star^2+M_3^2}\;,
\end{split}
\end{align}
respectively, and therefore we have $c_\rho^2=c_\theta^2 \equiv c_T^2$.
The deviation of the propagation speed of gravitational waves from that of light can be parameterized as
\begin{align}
\alpha_T(r) \equiv c_T^2-1=-\frac{M_3^2(r)}{M_\star^2+M_3^2(r)}\;,\label{def-alpha_T}
\end{align}
which represents the deviation from GR. Note that the LIGO/Virgo bound~\cite{TheLIGOScientific:2017qsa,GBM:2017lvd,Monitor:2017mdv} on $c_T^2$ implies that $|\alpha_T|\lesssim 10^{-15}$ in the asymptotic flat region.

Let us now write down the master equation for the odd modes, i.e., the generalized RW equation.
In doing so, it is useful to introduce the tortoise coordinate~$r_*$ for the odd mode via the following relation:
\begin{align}
\frac{\de r}{\de r_*} =\sqrt{\frac{B}{A}}\frac{A+\alpha_T}{\sqrt{1+\alpha_T}}\equiv F(r)\;. \label{def_F} 
\end{align}
The sound horizon~$r=r_g$ for the odd modes is determined by the condition~$F(r_g)=0$.
Then, from the action~(\ref{eq:chi_s}), the equation of motion for $\Psi \equiv (s_1 s_2)^{1/4} \chi$ is obtained as
\begin{align}
\left[\frac{\partial^2}{\partial r_*^2}-\frac{\partial^2}{\partial \tilde{t}^2}-F(r)V_\mathrm{eff}(r)\right]\Psi(\tilde{t},r_*)=0\;, \label{gene-RW}
\end{align}
where we have used the $(\tilde{t}, r)$ coordinates with\footnote{
The new time coordinate~$\tilde{t}$ has been introduced to remove a cross-derivative term~$\partial_t \partial_{r_*}\Psi$.
Note that a constant-$\tilde{t}$ hypersurface can be timelike in the vicinity of the odd-mode horizon.
Hence, when we perform a time-domain analysis in the presence of a matter source, the initial hypersurface should be chosen with care so that it is spacelike with respect to both the effective metric and the matter-frame metric.
(See \cite{Nakashi:2022wdg,Nakashi:2023vul} for related discussions.)
However, when we discuss TLNs in the subsequent Section, we only focus on static (or zero-frequency) perturbations, and therefore this problem is absent.
\label{footnote_ttilde}
}
\begin{align}
\tilde{t} = t+\int \de r\sqrt{\frac{1-A}{AB}}\frac{\alpha_T}{A+\alpha_T} \;, 
\end{align}
and the effective potential is given by
\begin{align}\label{eq:RW_potential}
V_{\rm eff}(r) 
&=\sqrt{1+\alpha_T}\sqrt{\frac{A}{B}}\frac{\ell(\ell+1)-2}{r^2}+\frac{r}{(1+\alpha_T)^{1/4}}\left[ F\cdot \left(\frac{(1+\alpha_T)^{1/4}}{r}\right)'\,\right].
\end{align}
Note that, in Eq.~(\ref{gene-RW}), we have factored out the function~$F(r)$ from the effective potential.
In the frequency space where $\Psi(\tilde{t}, r_*)=\tilde{\Psi}(r_*;\omega) e^{-i\omega \tilde{t}}$, we have
\begin{align}
\left[\frac{\de^2}{\de r_*^2}+\omega^2 -F(r)V_\mathrm{eff}(r)\right]\tilde{\Psi}(r_*; \omega)=0\;. \label{gene-RW-freq}
\end{align}
This is the main equation from which we are going to extract the TLNs in the next Section. 
We see that the differences from the case of GR are the definition of $r_*$, the time coordinate~$\tilde{t}$, and the profile of the effective potential.
In the case of the Schwarzschild solution in GR, Eq.~(\ref{gene-RW}) reduces to the standard RW equation.

\section{Analysis for tidal Love numbers based on the EFT approach}\label{sec:analysis}
\subsection{Tidal Love numbers of Schwarzschild BH}\label{sec:TLN_Schwarzschild}
\subsubsection{Definition of relativistic tidal Love numbers in GR}\label{sec:TLN_Schwarzschild-GR}
Tidal deformation is an important effect that can be observed in various astronomical scenarios, ranging from planetary scales to the merger of compact binary systems, not just in the context of binary BH mergers.
At the linear level, a tidal deformation induced by external static tidal fields is quantified by these TLNs, representing induced multipole moments. 
For instance, in the case of a binary system, the TLNs can measure the magnitude of tidal deformation of one body due to the force exerted from the other.

Let us now consider the standard RW equation in the static limit:
    \begin{align}
    \left[f(r)\frac{\de}{\de r}\left(f(r)\frac{\de}{\de r}\right) -f(r)V_\mathrm{RW}(r)\right]\tilde{\Psi}(r)=0\;, \qquad
    V_{\rm RW}(r) = \frac{\ell(\ell+1)}{r^2}-\frac{3r_g}{r^3}\;,
    \label{RW_GR}
    \end{align}
with $f(r) = 1-r_g/r$, for perturbations of multipole index~$\ell$.
Note that in this case $r_g$ coincides with the horizon for the background Schwarzschild metric, and $F(r) = f(r)$.
It is known that the general solution to Eq.~\eqref{RW_GR} can be written in terms of the hypergeometric functions, and it is straightforward to find the solution which is regular at the horizon.
The large-$r$ behavior of the solution is in general given by
    \begin{equation}
    \tilde{\Psi}(r)\sim \left(\frac{r}{r_g}\right)^{\ell+1}\left[1+\mathcal{O}\left(\frac{r_g}{r}\right)\right]+
    K_{\ell}\left(\frac{r}{r_g}\right)^{-\ell}\left[1+\mathcal{O}\left(\frac{r_g}{r}\right)\right],
    \label{TLN_def}
    \end{equation}
up to an overall constant, with $K_\ell$ being a constant.
Here, the first and second terms correspond to the growing and decaying modes as $r\to\infty$.
The growing mode is non-normalizable and should be regarded as an external field, while the decaying mode is understood as the response to it.
Therefore, the constant~$K_\ell$ characterizes the tidal response, and this corresponds to the TLN.\footnote{Some authors use different conventions/notations for TLNs. For instance, the TLNs are called $\kappa_{\ell}$ in \cite{Katagiri:2023yzm}, which is related to our $K_\ell$ by $K_\ell=[2(\ell+2)(\ell+1)/\ell(\ell-1)]\kappa_{\ell}$.}
Within the framework of GR, it is well established that the TLNs of Schwarzschild BH are vanishing~\cite{Damour:2009vw,Binnington:2009bb,Hui:2020xxx,LeTiec:2020bos,LeTiec:2020spy,Chia:2020yla,Charalambous:2021mea,Riva:2023rcm}.

However, when we define the TLNs with the asymptotic behavior of the static perturbation at spatial infinity, there is an ambiguity associated with the fact that there is no unique separation of the growing/decaying modes.
Indeed, since $\ell$ is an integer, the decaying mode in Eq.~\eqref{TLN_def} may be absorbed into the growing mode.
[Note that the factors of $1+{\cal O}(r_g/r)$ represent series expansions in $r_g/r$ starting with $(r_g/r)^0$.]
One of the possible ways to resolve this problem is to analytically continue $\ell$ to generic complex values, first introduced in \cite{Kol:2011vg} (see also \cite{Charalambous:2021mea,Creci:2021rkz}).\footnote{Also, if one considers dynamical perturbations, the effect of non-vanishing $\omega$ can be absorbed into a non-integer shift of $\ell$~\cite{Mano:1996mf}, and thus there is no such an ambiguity.
In this case, the static limit~$\omega\to 0$ corresponds to the limit where $\ell$ goes to an integer.
Other possible approaches include dimensional regularization~\cite{Kol:2011vg,Hui:2020xxx}, scattering theory~\cite{Creci:2021rkz,Katagiri:2023yzm}, and worldline EFT~\cite{Ivanov:2022hlo,Ivanov:2022qqt}.}
This allows us to separate the growing/decay modes and hence define the TLNs uniquely.

Before proceeding to a general analysis of TLNs based on our EFT, we shall discuss the simplest case of stealth Schwarzschild solution in the next Section.

\subsubsection{Stealth Schwarzschild solution in the EFT}\label{sec:Stealth-Master-eq}

In scalar-tensor theories (or any other modified gravity models), the solution for the metric is different from that in GR in general.
However, there is a special class of solutions where the metric has the same form as that in GR and the effect of a non-trivial scalar configuration is hidden, at least at the background level.
Such a solution is called a stealth solution~\cite{Mukohyama:2005rw,Motohashi:2018wdq,Takahashi:2020hso}, and perturbations about it were studied in, e.g.,~\cite{Babichev:2018uiw,Takahashi:2019oxz,deRham:2019gha,Motohashi:2019ymr,Khoury:2020aya,Tomikawa:2021pca,Takahashi:2021bml}.
In this Section, we show that the TLNs for the stealth Schwarzschild background are vanishing due to the fact that the generalized RW equation for this particular background is equivalent to that in GR. 
Note that the equivalence can be verified without setting $\omega=0$, and hence we keep $\omega$ non-zero in this Section.

For a stealth Schwarzschild solution, the background metric is given by
\begin{align}
A(r)=B(r)=1-\frac{r_s}{r}\;,
\end{align}
where $r_s$ is a constant that corresponds to the horizon radius.
In this case, the parameter~$\alpha_T$ [see Eq.~(\ref{def-alpha_T})] is a constant and the generalized RW equation~\eqref{gene-RW-freq} becomes
\begin{align}
\left[F(r) \frac{\de}{\de r}\left(F(r)\frac{\de}{\de r}\right)+\omega^2-F(r)V_\mathrm{eff}(r)\right]\tilde{\Psi}(r;\omega)=0\;, \quad
V_{\rm eff}(r)=\sqrt{1+\alpha_T}\left(\frac{\ell(\ell+1)}{r^2}-\frac{3r_g}{r^3}\right),
\label{genRW_stealth}
\end{align}
with $F(r)=\sqrt{1+\alpha_T}(1-r_g/r)$ and $r_g=r_s/(1+\alpha_T)$.
Interestingly, Eq.~\eqref{genRW_stealth} is equivalent to the RW equation in GR up to the following rescaling of $\omega$:
\begin{equation}
\begin{split}
\tilde{\omega}&=\frac{\omega}{\sqrt{1+\alpha_T}}
\;.
\end{split}\label{stealth-scaling}
\end{equation}
Indeed, the generalized RW equation now takes the form,
\begin{align}
\left[f(r) \frac{\de}{\de r} \left(f(r)\frac{\de}{\de r}\right) +\tilde{\omega}^2-f(r)V_\mathrm{RW}(r)\right]\tilde{\Psi}(r;\sqrt{1+\alpha_T}\,\tilde{\omega})=0\;, \quad
V_{\rm RW}(r)=\frac{\ell(\ell+1)}{r^2}-\frac{3r_g}{r^3}\;,
\end{align}
which is nothing but the RW equation in the GR case.
This suggests that the dynamics of the odd modes on a stealth Schwarzschild background is obtained from that in GR via a simple scaling~\cite{Mukohyama:2023xyf,Nakashi:2023vul}.
In particular, TLNs for a stealth Schwarzschild background are vanishing as in the GR case.
In other words, TLNs can be non-vanishing only if one considers a non-stealth background.

We note that the perturbations about a stealth Minkowski/de Sitter background (and hence in the asymptotic flat region of a stealth Schwarzschild background) have a vanishing sound speed and hence are strongly coupled~\cite{Motohashi:2019ymr}. A possible way out of this problem is to take into account a higher-derivative to introduce a $k^4$ term in the dispersion relation, which is known as the scordatura mechanism~\cite{Motohashi:2019ymr}. The effect of the scordatura term on the background spacetime was discussed in \cite{Mukohyama:2005rw,DeFelice:2022qaz}. 
Since the mode, which is strongly coupled, belongs to the even sector, one can safely use the generalized RW equation~\eqref{gene-RW} or \eqref{gene-RW-freq} to study the dynamics of the odd modes.

\subsection{Tidal Love numbers of non-stealth BH: methodology}
\label{ssec:TLN_extraction}

Let us now proceed to the analysis of TLNs for odd-parity perturbations about static and spherically symmetric BHs in our EFT framework.
For this purpose, we investigate the static solution for perturbations (i.e., $\omega=0$ in the frequency space) that is regular at the odd-mode horizon~$r=r_g$, based on the generalized RW equation~\eqref{gene-RW-freq}.\footnote{Another quantity that characterizes the tidal response for a low-frequency perturbation ($\omega\ll r_0^{-1}$, with $r_0$ being the radius of the object in concern) is the so-called dissipation number, which is associated with the dissipation of the tidal field at the event horizon. Imposing the ingoing boundary condition at the horizon in our setup, we have to care about the issue raised in Footnote~\ref{footnote_ttilde}.
(See \cite{Mano:1996vt,Mano:1996mf,Mano:1996gn} for analytic solutions to the Teukolsky/RW equation for $\omega\ne 0$ in GR.)}

\subsubsection{Perturbative solution}\label{ssec:pert_sol}

In what follows, we use the dimensionless radial coordinate~$x \equiv r/r_g$, so that the odd-mode horizon corresponds to $x=1$.
For static perturbations, the generalized RW equation~\eqref{gene-RW-freq} reads
\begin{align}
    \left[F(x)\frac{\de}{\de x}\left(F(x)\frac{\de}{\de x}\right)-F(x)\tilde{V}(x)\right]\tilde{\Psi}(x)=0\;, \label{eq:GRW-x}
\end{align}
where we have denoted $\tilde{\Psi}(r=r_g x,\omega=0)$ by $\tilde{\Psi}(x)$ and defined
\begin{align}
    \tilde{V}(x) \equiv r_g^2V_\mathrm{eff}(r)
    =\sqrt{1+\alpha_T}\sqrt{\frac{A}{B}}\frac{\ell(\ell+1)-2}{x^2}+\frac{x}{(1+\alpha_T)^{1/4}}\frac{\de}{\de x}\left[ F\cdot\frac{\de}{\de x}\left(\frac{(1+\alpha_T)^{1/4}}{x}\right)\right].\label{eq:GRW-potential-dimless}
\end{align}
It is useful to perform a change of variable so that all the modifications from GR are absorbed into the effective potential.
As shown in Appendix~B of \cite{Cardoso:2019mqo}, this can be achieved by introducing $\tilde{\psi}(x)=\sqrt{Z(x)}\,\tilde{\Psi}(x)$, with $Z(x)=F(x)/f(x)$ and $f(x)=1-1/x$.
Note that we assume $x=1$ is a single zero of $F(x)$, and hence the function~$Z(x)$ is regular at the horizon.
As a result, we have
\begin{align}
    &\left[f(x)\frac{\de}{\de x}\left(f(x)\frac{\de}{\de x}\right)-f(x)V(x)\right]\tilde{\psi}(x)=0\;,\label{eq:GRW-f-static}\\
    &V(x)\equiv \frac{\tilde{V}}{Z}-\frac{1}{4Z^2}\left[f\left(\frac{\de Z}{\de x}\right)^2-2Z\frac{\de}{\de x}\left(f\frac{\de Z}{\de x}\right)\right].\label{eq:GRW-potential}
\end{align}

Suppose that the effective potential~$V(x)$ in Eq.~\eqref{eq:GRW-f-static} can be expanded as 
    \begin{align}
        V(x)=\sum_{k\ge 0}\eta^k V^{(k)}(x)\;, \qquad
        V^{(0)}(x) = \frac{\ell(\ell+1)}{x^2}-\frac{3}{x^3}\;,
        \label{V_eta-expansion}
    \end{align}
with the parameter~$\eta$ representing the deviation from GR.\footnote{It is straightforward to extend our methodology to a more general situation where the deviation from GR is controlled by more than one parameter.}
Then, we can obtain the solution of Eq.~(\ref{eq:GRW-f-static}) perturbatively in $\eta$ by the Green's function method~\cite{DeLuca:2023mio,Riva:2023rcm}. 
Namely, we consider a solution of the form
\begin{align}
    \tilde{\psi}(x)=\sum_{k\ge 0}\eta^k \tilde{\psi}^{(k)}(x)\;.
    \label{psi_eta-expansion}
\end{align}
It should be noted that the leading asymptotic behavior of the solution remains the same as that in GR if $V^{(k)}(x)={\cal O}(x^{-3})$ for all $k\ge 1$ [see Eq.~\eqref{full-order_sol_1/x-expansion}].

Our strategy is to solve the differential equation~\eqref{eq:GRW-f-static} order by order in $\eta$.
At ${\cal O}(\eta^0)$, we obtain the standard RW equation:
\begin{align}
    \left[f(x)\frac{\de}{\de x}\left(f(x)\frac{\de}{\de x}\right)-f(x)V^{(0)}(x)\right]&\tilde{\psi}^{(0)}(x)=0\;.\label{eq:0-th-oDE}
\end{align}
The two independent solutions to this equation are written in terms of the hypergeometric function as
\begin{align}
    \tilde{\psi}_{\rm RW}^{+}(x)&=x^{\ell+1} {}_2F_1\left(-\ell-2,2-\ell;-2 \ell;\frac{1}{x}\right)
    =x^{\ell+1}\sum_{m=0}^{\infty} \frac{a_m^{+,(0)}(\ell)}{x^m} \;, \label{psiRW+} \\
    \tilde{\psi}_{\rm RW}^{-}(x)&= {x^{-\ell}} {}_2F_1\left(\ell-1,\ell+3;2 \ell+2;\frac{1}{x}\right)
    =x^{-\ell}\sum_{n=0}^\infty \frac{a_n^{-,(0)}(\ell)}{x^n} \;, \label{psiRW-}
\end{align}
where we have defined
\begin{align}
    a_m^{+,(0)}(\ell)&\equiv \frac{\Gamma(-2\ell)}{\Gamma(-\ell-2)\Gamma(2-\ell)}\frac{\Gamma(-\ell-2+m)\Gamma(2-\ell+m)}{m! \Gamma(-2\ell+m)}\;, \\
    a_n^{-,(0)}(\ell)&\equiv \frac{\Gamma(2\ell+2)}{\Gamma(\ell-1)\Gamma(\ell+3)}\frac{\Gamma(\ell-1+n)\Gamma(\ell+3+n)}{n! \Gamma(2\ell+2+n)}\;.
\end{align}
Here, the coefficients~$a_{0}^{\pm,(0)}(\ell)$ are normalized to unity and the Wronskian of the solutions is given by
    \begin{align}
    W\!\left[\tilde{\psi}_{\rm RW}^+(x),\tilde{\psi}_{\rm RW}^-(x)\right]
    \equiv f(x)\left(\tilde{\psi}_{\rm RW}^+(x)\frac{\de\tilde{\psi}_{\rm RW}^-(x)}{\de x}-\tilde{\psi}_{\rm RW}^-(x)\frac{\de\tilde{\psi}_{\rm RW}^+(x)}{\de x}\right)
    =-(2\ell+1)\;, \label{Wronskian}
    \end{align}
which is a constant.
The asymptotic behaviors of the solutions~(\ref{psiRW+}) and (\ref{psiRW-}) for large $x$ are given by
\begin{align}
    \tilde{\psi}_{\rm RW}^+(x)&\rightarrow
     x^{\ell+1} \left[1-\frac{(\ell-2) (\ell+2)}{2 \ell x}+\frac{(\ell-3) (\ell-2) (\ell+1) (\ell+2)}{4 \ell (2 \ell-1) x^2}+\mathcal{O}\left(x^{-3}\right)\right], \\
    \tilde{\psi}_{\rm RW}^-(x)&\rightarrow {x^{-\ell}} \left[1+\frac{(\ell-1) (\ell+3)}{(2 \ell+2) x}+\frac{(\ell-1) \ell (\ell+3) (\ell+4)}{4 (\ell+1) (2 \ell+3) x^2}+\mathcal{O}\left(x^{-3}\right)\right].
\end{align}
On the other hand, for generic $\ell\in\mathbb{C}$, both the solutions are divergent at the horizon.
Indeed, thanks to the expansion formula~\cite{Hui:2020xxx} around $z = 1$,
    \begin{align}
    {}_2F_1\left(a,b;a+b;z\right) \to -\frac{\Gamma(a+b)}{\Gamma(a)\Gamma(b)}\log (1-z)+\text{(finite terms)}\;,
    \end{align}
we have
    \begin{align}
    \tilde{\psi}_{\rm RW}^+(x)&\sim -\frac{\Gamma(-2\ell)}{\Gamma(-\ell-2)\Gamma(-\ell+2)}\log\left(1-\frac{1}{x}\right), \\
    \tilde{\psi}_{\rm RW}^-(x)&\sim -\frac{\Gamma(2\ell+2)}{\Gamma(\ell-1)\Gamma(\ell+3)}\log\left(1-\frac{1}{x}\right),
    \end{align}
in the vicinity of $x=1$.
Therefore, the horizon-regular solution of \eqref{eq:0-th-oDE} is written as
\begin{align}\label{eq:sol_RW_reg}
    \tilde{\psi}^{(0)}=\tilde{\psi}_{\rm RW}^{\text{hor-reg}}(x)=\tilde{\psi}_{\rm RW}^+(x)+K_\ell^{(0)}\tilde{\psi}_{\rm RW}^-(x)\;,
\end{align}
where $K_\ell^{(0)}$ is a constant given by
\begin{align}
    K_\ell^{(0)} = -\frac{\Gamma(-2\ell)}{\Gamma(-\ell-2)\Gamma(-\ell+2)}\frac{\Gamma(\ell-1)\Gamma(\ell+3)}{\Gamma(2\ell+2)}\;,
\end{align}
such that the divergence at $x = 1$ disappears.
For $\ell\in\mathbb{Z}_{\ge 2}$, corresponding to a physical multipole, we have $K_\ell^{(0)}=0$, which corresponds to the fact that the TLNs are vanishing in GR and the mode function~$\tilde{\psi}_{\rm RW}^+$ becomes purely a growing mode.\footnote{Precisely speaking, when we discuss $\ell\in\mathbb{Z}_{\ge 2}$, we consider a small deviation from $\ell$, say $\ell+\epsilon$ with $\epsilon\in\mathbb{C}$ and $|\epsilon|\ll 1$, and then take the limit~$\epsilon\to 0$. Here and in what follows, $\mathbb{Z}_{\ge n}$ denotes the set of integers that are greater than or equal to $n$.}

Let us now consider the differential equation~\eqref{eq:GRW-f-static} at ${\cal O}(\eta^k)$, which can be written as
\begin{equation}\label{eq:kth-order-eq}
    \left[f(x)\frac{\de}{\de x}\left(f(x)\frac{\de}{\de x}\right)-f(x)V^{(0)}(x)\right]\tilde{\psi}^{(k)}(x)=f(x)\mathcal{S}^{(k)}(x)
    \quad (k\ge 1) \;,
\end{equation}
where the source term~$\mathcal{S}^{(k)}$ is given by
\begin{equation}
     \mathcal{S}^{(k)}(x) \equiv \sum_{i=0}^{k-1} V^{(k-i)}(x)\tilde{\psi}^{(i)}(x)\;. \label{source_def}
\end{equation}
Note that in Eq.~(\ref{eq:kth-order-eq}) the derivative operator acting on $\tilde{\psi}^{(k)}$ is the same as that in the zeroth-order equation~\eqref{eq:0-th-oDE}.
Therefore, the homogeneous solution is given by a linear combination of $\tilde{\psi}_{\rm RW}^{\pm}(x)$ defined in Eqs.~\eqref{psiRW+} and \eqref{psiRW-}.
Given the solution~$\tilde{\psi}^{(i)}$ for $i=0,1,\cdots,k-1$, then $\tilde{\psi}^{(k)}$ can be perturbatively obtained by solving Eq.~\eqref{eq:kth-order-eq}.
For this purpose, we introduce the Green's function~$G(x,x')$ which satisfies
\begin{align}
    \left[f(x)\frac{\de}{\de x}\left(f(x)\frac{\de}{\de x}\right)-f(x)V^{(0)}(x)\right]G(x,x')=f(x)\delta(x-x')\;, \label{GreenF_def}
\end{align}
with $\delta(x)$ being the delta function.
We impose the boundary conditions that the field is regular at $x=1$ and $\tilde{\psi}^{(k)}(x)={\cal O}(x^{\ell})$ as $x\to\infty$, where the latter comes from the fact that one can always renormalize the full-order solution~$\tilde{\psi}(x)$ so that the coefficient of $x^{\ell+1}$ is unity, i.e., the correction is of ${\cal O}(x^{\ell})$.
Then, the Green's function is
\begin{align}
    G(x,y)=-\frac{1}{2\ell+1}\left[\tilde{\psi}_{\rm RW}^-(x)\tilde{\psi}_{\rm RW}^{\text{hor-reg}}(y)\Theta(x-y)+\tilde{\psi}_{\rm RW}^{\text{hor-reg}}(x)\tilde{\psi}_{\rm RW}^-(y)\Theta(y-x)\right], 
    \label{Green_func}
\end{align}
where $\Theta(x)$ is the Heaviside step function.
The detailed derivation of the Green's function is provided in Appendix~\ref{app:detail-Green}.
As a result, the inhomogeneous solution of \eqref{eq:kth-order-eq} is given by
\begin{align}
    \tilde{\psi}^{(k)}(x)&=\int_1^\infty \de y\  G(x,y)\mathcal{S}^{(k)}(y) \nonumber \\
    &=-\frac{1}{2\ell+1}\left[\tilde{\psi}_{\rm RW}^-(x)\int_1^x \de y\,\tilde{\psi}_{\rm RW}^{\text{hor-reg}}(y)\mathcal{S}^{(k)}(y)
    -\tilde{\psi}_{\rm RW}^{\text{hor-reg}}(x)\int_\infty^x \de y\,\tilde{\psi}_{\rm RW}^-(y)\mathcal{S}^{(k)}(y)\right].
    \label{k-th-order-sol-0}
\end{align}
Alternatively, by using the relation~$\tilde{\psi}_{\rm RW}^{\text{hor-reg}}(x)=\tilde{\psi}_{\rm RW}^+(x)+K_\ell^{(0)}\tilde{\psi}_{\rm RW}^-(x)$, the solution~(\ref{k-th-order-sol-0}) becomes 
\begin{align}
    \tilde{\psi}^{(k)}(x)
    =-\frac{1}{2\ell+1}\bigg[&\tilde{\psi}_{\rm RW}^-(x)\int_1^x\de y\,\tilde{\psi}_{\rm RW}^+(y)\mathcal{S}^{(k)}(y)-\tilde{\psi}_{\rm RW}^+(x)\int_1^x \de y\,\tilde{\psi}_{\rm RW}^-(y)\mathcal{S}^{(k)}(y) \nonumber \\
    &+
    \left(\tilde{\psi}_{\rm RW}^+(x)+K_\ell^{(0)}\tilde{\psi}_{\rm RW}^-(x)\right)\int_1^\infty \de y\,\tilde{\psi}_{\rm RW}^-(y)\mathcal{S}^{(k)}(y)\bigg]\;.
    \label{k-th-order-sol}
\end{align}
In the next Section, we will study the asymptotic behavior of each term in the solution above in order to extract the static TLNs.
Notice that so far we have not specified the explicit form of the effective potential.
Therefore, our result in this Section generically applies to arbitrary static and spherically symmetric BHs, as long as the deviations from GR can be treated using perturbation theory.

\subsubsection{Extraction of tidal Love numbers}\label{ssec:extraction_TLN}

In the previous Section, we have obtained the solution to the differential equation~\eqref{eq:GRW-f-static} order by order in $\eta$.
Here, we expand the solution in $1/x$ to discuss the behavior of the solution at spatial infinity and how to extract TLNs.
As mentioned earlier, we now assume that the correction to the effective potential is $V^{(k)}(x)={\cal O}(x^{-3})$ for $k\ge 1$, and therefore the behavior of the full-order solution at large $x$ is given by
    \begin{align}
    \tilde{\psi}(x)=x^{\ell+1}\left[1+{\cal O}(x^{-1})\right]+K_\ell(\eta)x^{-\ell}\left[1+{\cal O}(x^{-1})\right],
    \label{full-order_sol_1/x-expansion}
    \end{align}
up to an overall constant,\footnote{Note that we are interested in the ratio of the induced multipole moment to the tidal multipole moment, and the result is invariant under a multiplication of an overall factor.
We have used this degree of freedom to normalize the field~$\tilde{\psi}$, so that the coefficient in front of $x^{\ell+1}$ is unity.}
with
    \begin{align}
    K_\ell(\eta)=\sum_{k\ge 0}\eta^k K_\ell^{(k)}\;,
    \end{align}
which corresponds to the TLNs involving the correction of ${\cal O}(\eta^k)$ ($k\ge 1$).
It should be noted that there is no overlap between the series expansion starting with $x^{\ell+1}$ and that starting with $x^{-\ell}$ for a generic $\ell\in\mathbb{C}$.
Note also that the TLNs do not depend on the choice of the master variable.
For instance, we can go back to the original master variable~$\tilde{\Psi}=Z(x)^{-1/2}\tilde{\psi}(x)$ in Eq.~\eqref{eq:GRW-x}, and this does not mix the terms of $x^{\ell+1}$ and $x^{-\ell}$ in Eq.~\eqref{full-order_sol_1/x-expansion} so long as $Z(x)$ is analytic everywhere on or outside the horizon and it goes to a constant as $x\to\infty$.
Therefore, up to an overall constant, $\tilde{\Psi}$ has the same large-$x$ behavior as that of $\tilde{\psi}$.
Moreover, the metric perturbations [$h_0$ and $h_1$ in Eq.~\eqref{eq:odd_pert}] can be reconstructed from the master variable $\tilde{\psi}$,\footnote{More precisely, one first uses $\tilde{\Psi}=Z(x)^{-1/2}\tilde{\psi}(x)$ to obtain $\tilde{\Psi}$. Then, using $\Psi = \tilde{\Psi} e^{-i\omega \tilde{t}}$ and $\chi = (s_1 s_2)^{-1/4} \Psi$, one recovers $\chi$. Finally, the perturbations~$h_0$ and $h_1$ are obtained using Eq.~(5.10) of \cite{Mukohyama:2022skk}.} and the TLNs that can be read off from them coincide with $K_\ell$ in Eq.~\eqref{full-order_sol_1/x-expansion}.

Let us now define the $1/x$-expansion of the ${\cal O}(\eta^k)$ coefficients in Eqs.~\eqref{V_eta-expansion} and \eqref{psi_eta-expansion} as
    \begin{align}
    V^{(k)}(x)&=\sum_{j=3}^{N_{k}}\frac{v_{j}^{(k)}(\ell)}{x^j}
    \quad (k\ge 1)\;, \label{Vk_1/x-expansion} \\
    \tilde{\psi}^{(k)}(x)&=x^{\ell+1}\sum_{p\ge 0} \frac{a_p^{+,(k)}(\ell)}{x^p}+K_\ell^{(k)}x^{-\ell}\sum_{q\ge 0}\frac{a_q^{-,(k)}(\ell)}{x^q}
    \quad (k\ge 0)\;. \label{psik_1/x-expansion}
    \end{align}
For $k\ge 1$, we have $a_0^{+,(k)}=0$ due to the boundary condition imposed on the Green's function (see discussions in the previous Section).
Note that we have assumed that the expansion stops at a finite order~$N_k$ for $V^{(k)}(x)$ with $k\ge 1$, which is the case for the Hayward background studied in Section~\ref{ssec:Hayward}.
Then, the source term~${\cal S}^{(k)}(x)$ in Eq.~\eqref{source_def} can be expressed as
    \begin{align}
    {\cal S}^{(k)}(x)=\mathcal{S}^{(k)}_+(x)+\mathcal{S}^{(k)}_-(x)\;,
    \end{align}
with
    \begin{equation}
    \begin{split}\label{eq:Source-split}
    \mathcal{S}^{(k)}_+(x)&\equiv \sum_{i=0}^{k-1}\sum_{j=3}^{N_{k-i}}v_{j}^{(k-i)}x^{\ell+1-j}\sum_{p\ge 0} \frac{a_p^{+,(i)}}{x^p}\;, \\
    \mathcal{S}^{(k)}_-(x)&\equiv \sum_{i=0}^{k-1}K_\ell^{(i)}\sum_{j=3}^{N_{k-i}}v_{j}^{(k-i)}x^{-\ell-j}\sum_{q\ge 0}\frac{a_q^{-,(i)}}{x^q}\;.
    \end{split}
    \end{equation}
When we obtain the perturbative solution in Eq.~\eqref{k-th-order-sol}, we need to compute integrals of the form
    \begin{align}
    I^{(k)}_{ss'}(x)
    \equiv \int_{1}^{x}\de y\,\tilde{\psi}_{\rm RW}^{s}(y)\mathcal{S}_{s'}^{(k)}(y)\;,
        \label{Iss'_def}
    \end{align}
with $s,s'\in\{+,-\}$.
A term-by-term integration yields\footnote{In Eq.~\eqref{Iss'_def}, the integrand~$\tilde{\psi}_{\rm RW}^{s}(y) {\cal S}_{s'}^{(k)}(y)$ is regular except at the horizon~$y=1$, and it diverges at $y=1$. However, the divergence at $y=1$ is only as fast as $[\log(y-1)]^2$, which is integrable in the vicinity of $y=1$. This means that one can choose a function that is integrable in the domain~$[1,x]$ ($x<\infty$) and that dominates every partial sum of the series expansion of the integrand. Therefore, one can apply the dominated convergence theorem~\cite{rudin1976principles} to justify the term-by-term integration.}
\begin{align}
    I^{(k)}_{++}(x)
    &=\sum_{i=0}^{k-1}\sum_{j=3}^{N_{k-i}}{v_{j}^{(k-i)}}\int_1^x \de y\,y^{\ell+1-j}\left(\sum_{p\ge 0} \frac{a_p^{+, (i)}}{y^p}\right) y^{\ell+1}\sum_{m\ge 0}\frac{a_m^{+,(0)}}{y^m}\nonumber\\
    &=C^{(k)}_{++}+\sum_{i=0}^{k-1}\sum_{j=3}^{N_{k-i}}\sum_{p\ge 0}\sum_{m\ge 0}
    {v_{j}^{(k-i)}}a_p^{+,(i)}{a_m^{+,(0)}}\frac{x^{2\ell-j-m-p+3}}{2\ell-j-m-p+3}\;,\label{I++}\\
    I^{(k)}_{-+}(x)
    &=\sum_{i=0}^{k-1}\sum_{j=3}^{N_{k-i}}{v_{j}^{(k-i)}}\int_1^x \de y\,y^{\ell+1-j}\left(\sum_{p\ge 0} \frac{a_p^{+, (i)}}{y^p}\right) y^{-\ell}\sum_{n\ge 0}\frac{a_n^{-,(0)}}{y^n}\nonumber\\
    &=C^{(k)}_{-+}+\sum_{i=0}^{k-1}\sum_{j=3}^{N_{k-i}}\sum_{p\ge 0}\sum_{n\ge 0}{v_{j}^{(k-i)}}a_p^{+,(i)}{a_n^{-,(0)}}
    \frac{x^{-j-n-p+2}}{-j-n-p+2}\;,\\
    I^{(k)}_{+-}(x)
    &=\sum_{i=0}^{k-1}K_\ell^{(i)}\sum_{j=3}^{N_{k-i}}{v_{j}^{(k-i)}}\int_1^x \de y\,y^{-\ell-j}\left(\sum_{q\ge 0} \frac{a_q^{-, (i)}}{y^q}\right) y^{\ell+1}\sum_{m\ge 0}\frac{a_m^{+,(0)}}{y^m}\nonumber\\
    &=C^{(k)}_{+-}+\sum_{i=0}^{k-1}\sum_{j=3}^{N_{k-i}}\sum_{q\ge 0}\sum_{m\ge 0}
    K_\ell^{(i)}{v_{j}^{(k-i)}}a_q^{-,(i)}a_m^{+,(0)}\frac{x^{-j-m-q+2}}{-j-m-q+2}\;,\\
    I^{(k)}_{--}(x)
    &=\sum_{i=0}^{k-1}K_\ell^{(i)}\sum_{j=3}^{N_{k-i}}{v_{j}^{(k-i)}}\int_1^x \de y\,y^{-\ell-j}\left(\sum_{q\ge 0} \frac{a_q^{+, (i)}}{y^q}\right) y^{-\ell}\sum_{n\ge 0}\frac{a_n^{-,(0)}}{y^n}\nonumber\\
    &=C^{(k)}_{--}+\sum_{i=0}^{k-1}\sum_{j=3}^{N_{k-i}}\sum_{q\ge 0}\sum_{n\ge 0}
    K_\ell^{(i)}{v_{j}^{(k-i)}}a_q^{-,(i)}a_n^{-,(0)}\frac{x^{-2\ell-j-n-q+1}}{-2\ell-j-n-q+1}\;,
\end{align}
where $C^{(k)}_{ss'}$ being constants fixed so that $I^{(k)}_{ss'}(1)=0$.
Moreover, we define
    \begin{align}
    \tilde{I}^{(k)}_{ss'} \equiv I^{(k)}_{ss'}-C^{(k)}_{ss'}\;,
    \end{align}
so that
    \begin{align}
    \begin{split}
    &\tilde{I}^{(k)}_{++}\sim x^{2\ell}\left[1+{\cal O}(x^{-1})\right], \qquad
    \tilde{I}^{(k)}_{-+}\sim x^{-1}\left[1+{\cal O}(x^{-1})\right], \\
    &\tilde{I}^{(k)}_{+-}\sim x^{-1}\left[1+{\cal O}(x^{-1})\right], \qquad
    \tilde{I}^{(k)}_{--}\sim x^{-2\ell-2}\left[1+{\cal O}(x^{-1})\right].
    \end{split}
    \end{align}
Note that we have implicitly assumed that $\ell$ is a generic complex number, and therefore the denominator in Eq.~\eqref{I++} is non-vanishing.
If the denominator vanishes for some triplets~$(j,p,m)$ when $\ell\in\mathbb{Z}_{\ge 2}$, one has to treat them separately, as they lead to a logarithmic term (unless the coefficient~$v_{j}^{(k-i)}a_p^{+,(i)}$ is non-vanishing).
We discuss such a case separately later and for now assume that the above formulae for $I^{(k)}_{ss'}$ make sense even for $\ell\in\mathbb{Z}_{\ge 2}$.
In terms of $I^{(k)}_{ss'}$, the solution~$\tilde{\psi}^{(k)}$ in Eq.~\eqref{k-th-order-sol} can be written as
\begin{align}
    -(2\ell+1)\tilde{\psi}^{(k)}(x)&=\left[\tilde{I}^{(k)}_{++}(x)\tilde{\psi}_{\rm RW}^-(x)-\tilde{I}^{(k)}_{-+}(x)\tilde{\psi}_{\rm RW}^+(x)\right] \nonumber \\
    &\quad +\left[\tilde{I}^{(k)}_{+-}(x)\tilde{\psi}_{\rm RW}^-(x)-\tilde{I}^{(k)}_{--}(x)\tilde{\psi}_{\rm RW}^+(x)\right] \nonumber \\
    &\quad +\left[C^{(k)}_{++}+C^{(k)}_{+-}+K_\ell^{(0)}\left(C^{(k)}_{-+}+C^{(k)}_{--}\right)\right]\tilde{\psi}_{\rm RW}^-(x)\;. \label{eq:inhomo-generic}
\end{align}
Since $\tilde{\psi}_{\rm RW}^+(x)\sim x^{\ell+1}[1+{\cal O}(x^{-1})]$ and $\tilde{\psi}_{\rm RW}^-(x)\sim x^{-\ell}[1+{\cal O}(x^{-1})]$ as $x\to\infty$, the first line of Eq.~\eqref{eq:inhomo-generic} behaves as
    \begin{align}
    \tilde{I}^{(k)}_{++}(x)\tilde{\psi}_{\rm RW}^-(x)-\tilde{I}^{(k)}_{-+}(x)\tilde{\psi}_{\rm RW}^+(x)
    \sim x^{\ell}\left[1+{\cal O}(x^{-1})\right],
    \label{subleading_growing}
    \end{align}
and therefore only contributes to subleading terms of the growing mode.
Likewise, the second line of Eq.~\eqref{eq:inhomo-generic} behaves as
    \begin{align}
    \tilde{I}^{(k)}_{+-}(x)\tilde{\psi}_{\rm RW}^-(x)-\tilde{I}^{(k)}_{--}(x)\tilde{\psi}_{\rm RW}^+(x)
    \sim x^{-\ell-1}\left[1+{\cal O}(x^{-1})\right],
    \label{subleading_decaying}
    \end{align}
as $x\to\infty$, which is a subleading contribution to the decaying mode.\footnote{Equations~\eqref{subleading_growing} and \eqref{subleading_decaying} can be understood as corrections due to the modification of gravity in a region far from the black hole horizon.}
Hence, the large-$x$ behavior of $\tilde{\psi}^{(k)}$ is given by 
    \begin{align}\label{eq:psi_k_pert}
    \tilde{\psi}^{(k)}(x) = {\cal C}x^{\ell}\left[1+{\cal O}(x^{-1})\right]-\frac{1}{2\ell+1}\left[C^{(k)}_{++}+C^{(k)}_{+-}+K_\ell^{(0)}\left(C^{(k)}_{-+}+C^{(k)}_{--}\right)\right]x^{-\ell}\left[1+{\cal O}(x^{-1})\right],
    \end{align}
with ${\cal C}$ being a constant.
Notice that the leading corrections to the growing mode goes as $x^\ell$, since the coefficient of $x^{\ell + 1}$ has been normalized to unity [see discussion below Eq.~(\ref{GreenF_def})].
Comparing this with Eq.~\eqref{full-order_sol_1/x-expansion}, the contribution to the TLN of $\mathcal{O}(\eta^k)$ can be read off as
\begin{align}
K_\ell^{(k)}=-\frac{1}{2\ell+1}\left[C^{(k)}_{++}+C^{(k)}_{+-}+K_\ell^{(0)}\left(C^{(k)}_{-+}+C^{(k)}_{--}\right)\right]. \label{TLN-generic}
\end{align}
For $\ell\in\mathbb{Z}_{\ge 2}$, we have $K_\ell^{(0)}=0$, and hence only the first two terms in Eq.~\eqref{TLN-generic} contribute to $K_\ell^{(k)}$.
Therefore, one can read off $K_\ell^{(k)}$ from the constant part of the following integral:
    \begin{align}
    I[{\cal S}^{(k)}] =
    -\frac{1}{2\ell+1}\left[I^{(k)}_{++}(x)+I^{(k)}_{+-}(x)\right]
    =-\frac{1}{2\ell+1}\int_1^x \de y\,\tilde{\psi}_{\rm RW}^+(y)\mathcal{S}^{(k)}(y)\;.
    \label{int_TLN}
    \end{align}

Let us now discuss the logarithmic term that can arise when $\ell\in\mathbb{Z}_{\ge 2}$.
As mentioned earlier, this happens when the denominator of Eq.~\eqref{I++} vanishes.
The logarithmic term can be given by
    \begin{align}
    I^{(k)}_{++}(x)
    \supset \sum_{i=0}^{k-1}\sum_{j,p,m}v_{j}^{(k-i)}a_p^{+,(i)}a_m^{+,(0)}\log x\;,
    \label{I++_log}
    \end{align}
where the summation on $j,p,m$ runs over all the triplets~$(j,p,m)$ that satisfy
    \begin{align}
    3\le j\le N_{k-i}\;, \qquad
    p\ge 0\;, \qquad
    m\ge 0\;, \qquad
    2\ell-j-m-p+3=0\;. \label{sum_jpm}
    \end{align}
If such a logarithmic term shows up in $I^{(k)}_{++}(x)$, the large-$x$ expansion~\eqref{full-order_sol_1/x-expansion} of the full-order solution acquires a logarithmic correction to the $x^{-\ell}$ term:
    \begin{align}
    \tilde{\psi}(x)=x^{\ell+1}\left[1+{\cal O}(x^{-1})\right]+K_\ell^{\log}(\eta)x^{-\ell}\left[\log x+{\cal O}(x^0)\right],
    \end{align}
where
    \begin{align}
    K_\ell^{\log}(\eta)=\sum_{k\ge 1}\eta^k K_\ell^{(k),\log}\;.
    \end{align}
This corresponds to the so-called logarithmic running of TLNs (see, e.g.,~\cite{Cardoso:2019rvt,Hui:2020xxx}).
More explicitly, from Eq.~\eqref{I++_log}, we have
    \begin{align}
    K_\ell^{(k),\log}=-\frac{1}{2\ell+1}\sum_{i=0}^{k-1}\sum_{j,p,m}v_{j}^{(k-i)}a_p^{+,(i)}a_m^{+,(0)}\;,
    \end{align}
where again the range of the summation on $j,p,m$ is defined by Eq.~\eqref{sum_jpm}.
One can read off $K_\ell^{(k),\log}$ directly from the integral~$I[{\cal S}^{(k)}]$ as a coefficient of $\log x$ in the large-$x$ expansion.
Note that only $I^{(k)}_{++}(x)$ in $I[{\cal S}^{(k)}]$ can generate a logarithmic term.

To summarize, what we have learned in this Section is that the TLN (or its logarithmic running) of ${\cal O}(\eta^k)$ can be read off from the constant (or logarithmic) part of the integral~$I[{\cal S}^{(k)}]$ defined in Eq.~\eqref{int_TLN}.
In principle, our methodology can be applied to the study of TLNs associated with the odd modes for a large class of static and spherically symmetric BHs within our EFT.

\subsection{Example: Hayward background}\label{ssec:Hayward}

As an application of our general methodology developed in the previous Section, we consider the Hayward BH~\cite{Hayward:2005gi} as the background metric.\footnote{It is known that the Hayward BH does not satisfy some of the energy conditions at least in a portion of the spacetime~\cite{Maeda:2021jdc}, but this does not lead to an immediate problem in the present context.}
The Hayward BH is one of the best-known examples of non-singular BH, and such a BH would be interesting in the context of information loss problem~\cite{Frolov:2014jva}.
In particular, it was shown that the Hayward metric can be realized as an exact solution in DHOST theories~\cite{Babichev:2020qpr}, as well as in our EFT~\cite{Mukohyama:2023xyf}.

The metric for the Hayward BH solution is given by
\begin{align}
A(r)=B(r)=1-\frac{r_s r^2}{r^3+\sigma^3}\;, \label{eq:Hayward-metric}
\end{align}
where $r_s$ and $\sigma$ are constants of length dimension.
Note that $r_s$ corresponds to twice the ADM mass (times the gravitational constant).
The metric asymptotes to the Schwarzschild metric at large $r$, while it has a de Sitter core near the center ($r=0$) in the case of $\sigma>0$, and hence the curvature singularity at $r=0$ is resolved.\footnote{Rigorously speaking, the geometry at $r=0$ is singular since the Misner-Sharp mass is not even with respect to $r$. This is not a problem for our purpose in the present paper as we are interested in the behavior of the geometry on and outside the odd-mode horizon.}
Having said that, in the present paper, we regard $\sigma$ as just a phenomenological parameter that characterizes the deviation from the Schwarzschild BH on and outside the horizon, and we also consider $\sigma<0$.

For the Hayward metric, the function~$F(r)$ in Eq.~\eqref{def_F} is given by
\begin{align}
F(r)&=\frac{r^4-r_s(r^3+\sigma^3)}{r(r^3+\sigma^3)}\;.
\end{align}
We recall that the position of the odd-mode horizon~$r_g$ is determined by $F(r_g) = 0$.
Also, we assume $\sigma^3/r_s^3>-27/256$ so that the effective metric for the odd mode has a non-degenerate horizon~\cite{Mukohyama:2023xyf}.
For $\sigma<0$, there are two distinct zeros of $F(r)=0$ in the region~$r>0$, and the larger one is identified as $r=r_g$.

Written in terms of $x \equiv r/r_g$ and $\eta \equiv \sigma^3/r_g^3$, the functions~$A(r)$, $\alpha_T(r)$, and $F(r)$ take the form
\begin{equation}
    \begin{split}
A(x) &= 1 - \frac{x^2}{(1+\eta)(x^3+\eta)}\;, \quad 
\alpha_T(x) = -\frac{\eta(2x^3+\eta)}{(x^3+\eta)^2}\;,\\
F(x) &= \left(1-\frac{1}{x}\right)\frac{(1+\eta)x^3+\eta x^2+\eta x+\eta}{(1+\eta)(x^3+\eta)}\;.
\end{split}
\end{equation}
Note that $r_g=r_s(1+\eta)$ and the condition~$\sigma^3/r_s^3>-27/256$ corresponds to $\eta>-1/4$.
The effective potential~$V(x)$ in Eq.~\eqref{eq:GRW-f-static} can be written as
\begin{equation}
V(x)=\frac{\ell(\ell+1)(1+\eta)x}{x^3+\eta(x+1)(x^2+1)}+\frac{\sum_{i=0}^{12}V_i(\eta) x^i}{4 x^3 \left(x^3+\eta\right)^2 \left[x^3+\eta(x+1)(x^2+1)\right]^2}\;, \label{V-hayward-full}
\end{equation}
where
\begin{equation}\label{V-hayward-full-term}
    \begin{split}
         &V_0=3\eta^4\;, \quad
         V_1=5\eta^4\;, \quad
         V_2=6\eta^4\;,\quad V_3=6\eta^3(4+\eta)\;, \quad
         V_4=3\eta ^3 (14+5\eta)\;, \\
         &V_5=3\eta ^3 (14+3\eta)\;, \quad
         V_6=\eta ^2 \left(27+40\eta+4\eta^2\right), \quad
         V_7=-3\eta ^2(13+14\eta)\;, \\
         &V_8=-6 \eta ^2 (7+9\eta)\;, \quad
         V_9=-2 \eta \left(3+23\eta+32\eta^2\right), \quad
         V_{10}=-\eta \left(76+129\eta+72\eta ^2\right), \\
         &V_{11}=-3\eta (2-3\eta)\;, \quad
         V_{12}=-4 \left(3+2\eta-\eta^2\right).
    \end{split}
\end{equation}
Therefore, the first few expansion coefficients in Eq.~\eqref{V_eta-expansion} are given by
    \begin{align}
    \begin{split}
    V^{(1)}(x)&=-\frac{\ell(\ell+1)(x^2+x+1)}{x^5}+\frac{8x^3+9x^2-26x+21}{2x^6}\;, \\
    V^{(2)}(x)&=\frac{\ell(\ell+1)(x+1)(x^2+1)(x^2+x+1)}{x^8} \\
    &\quad -\frac{16x^6+35x^5+57x^4+30x^3+10x^2-157x+69}{x^9}\;.
    \end{split}\label{V_Hayward_eta_exp}
    \end{align}
Recall that $V^{(0)}(x)$ is nothing but the RW potential.
One can show that the $1/x$-expansion of $V^{(k)}$ has the form
\begin{align}
V^{(k)}(x)=\sum_{j=3}^{3(1+k)}\frac{v_j^{(k)}(\ell)}{x^j}
\quad (k\ge 1) \;.
\end{align}

In what follows, we demonstrate the computation of the TLNs for the Hayward background for $\ell=2,3,4$, based on our methodology developed in Section~\ref{ssec:TLN_extraction}.\footnote{One can alternatively apply the parameterized TLN formalism~\cite{Katagiri:2023umb} to obtain the TLN up to the first order in $\eta$, which we will discuss in Appendix~\ref{app:pTLN}.}
It is useful to note that the two independent solutions to the RW equation, Eqs.~\eqref{psiRW+} and \eqref{psiRW-}, can be written in terms of elementary functions for $\ell\in\mathbb{Z}_{\ge 2}$.
Written explicitly,
    \begin{align}
    \tilde{\psi}_{\rm RW}^+(x)&=x^{\ell+1}\sum_{n=0}^{\ell-2} \frac{a_n^+(\ell)}{x^n}\;, \quad
    \tilde{\psi}_{\rm RW}^-(x)={\cal B}(\ell)\left[\tilde{\psi}_{\rm RW}^+(x)\log \left(1-\frac{1}{x}\right)+x^{\ell+1} \sum_{n=1}^{\ell+2} \frac{b_n^-(\ell)}{x^n}\right],
    \label{psi+}
    \end{align}
where 
\begin{equation}
\begin{split}
    a_n^+(\ell)&\equiv \frac{(-\ell-2)_n (2-\ell)_n}{n!(-2\ell)_n}\;, \quad 
    b_n^-(\ell)\equiv \sum_{m=1}^{\ell+2}\frac{a_{n-m}^+(\ell)}{m}\;,\quad 
    {\cal B}(\ell) \equiv -(2\ell+1)\binom{2\ell}{\ell-2}^2\;,
    \label{calB}
\end{split}
\end{equation}
with $(c)_n$ denoting the rising factorial defined by $(c)_n\equiv c(c+1)\cdots(c+n-1)$.

\subsubsection{\texorpdfstring{$\ell=2$ case}{l=2 case}}\label{sec:Hayward-l=2}
In the case of $\ell=2$, we have
\begin{align}
    \tilde{\psi}_{\rm RW}^+(x)&=x^3=\tilde{\psi}_{\ell=2}^{(0)}(x)\;,  \label{psi+_l=2} \\
   \tilde{\psi}_{\rm RW}^-(x)&=-5\left[x^3 \log \left(1-\frac{1}{x}\right)+x^2+\frac{x}{2}+\frac{1}{3}+\frac{1}{4 x}\right].  \label{psi-_l=2}
\end{align}
Note that $\tilde{\psi}_{\rm RW}^-(x)={\cal O}(x^{-2})$ as $x\to\infty$.
We compute the solution to the differential equation~\eqref{eq:kth-order-eq} as well as the TLNs order by order in $\eta$, by use of the Green's function~\eqref{Green_func} (with $\tilde{\psi}_{\rm RW}^{\text{hor-reg}}=\tilde{\psi}_{\rm RW}^+$).
As we shall see in Appendix~\ref{app:exact-solutions}, for $\ell=2$, an exact solution to Eq.~\eqref{eq:GRW-f-static} as well as the TLNs can be obtained without resorting to the expansion in $\eta$, and the following results can be reproduced from the exact expressions.

\begin{itemize}
\item {\it First order in $\eta$}.
The effective potential and the source term at $\mathcal{O}(\eta)$ are given by
\begin{align}
V^{(1)}(x)&=-\frac{2}{x^3}-\frac{3}{2 x^4}-\frac{19}{x^5}+\frac{21}{2 x^6}\;, \\
\mathcal{S}^{(1)}(x)&=V^{(1)}(x)\tilde{\psi}_{\rm RW}^+(x)=
-2-\frac{3}{2 x}-\frac{19}{x^2}+\frac{21}{2 x^3}\;,
\end{align}
from which we obtain the first-order solution as
\begin{align}
\tilde{\psi}_{\ell=2}^{(1)}(x)&=\int_1^\infty \de y\,G(x,y)\mathcal{S}^{(1)}(y)=\frac{x^2}{2}+\frac{x}{2}+\frac{7}{2}\;. \label{l=2-1st-order}
\end{align}
Recall that the Green's function has been constructed so that the solution is regular at $x=1$ and $\tilde{\psi}_{\ell}^{(k)}={\cal O}(x^{\ell})$ as $x\to\infty$.
Then, the integral~\eqref{int_TLN} can be computed as
    \begin{align}
    I[{\cal S}^{(1)}]
    =-\frac{1}{5}\int_1^x \de y\,\tilde{\psi}_{\rm RW}^+(y)\mathcal{S}^{(1)}(y)
    =\frac{x^4}{10}+\frac{x^3}{10}+\frac{19}{10x^2}-\frac{21x}{10}\;,
    \end{align}
which does not contain a constant term.
Therefore, we obtain
    \begin{align}
    K_{\ell=2}^{(1)}=0\;.
    \end{align}

\item {\it Second order in $\eta$}.
The effective potential and the source term of $\mathcal{O}(\eta^2)$ are given by
\begin{align}
    V^{(2)}(x)&=
    {\frac{2}{x^3}+\frac{13}{4 x^4}+\frac{15}{4 x^5}+\frac{21}{2 x^6}+\frac{19}{2 x^7}+\frac{181}{4 x^8}-\frac{69}{4 x^9}}\;, \\
    \mathcal{S}^{(2)}(x)&=V^{(2)}(x)\tilde{\psi}_{\rm RW}^+(x)+V^{(1)}(x)\tilde{\psi}^{(1)}(x)\nonumber\\
    &=
    2+\frac{9}{4 x}+\frac{2}{x^2}-\frac{27}{4 x^3}-\frac{16}{x^5}+\frac{39}{2 x^6}\;, \label{l=2:2nd-order-source}
\end{align}
from which we obtain the second-order solution as
\begin{align}
    \tilde{\psi}_{\ell=2}^{(2)}(x)&=\int_1^\infty \de y\,G(x,y)\mathcal{S}^{(2)}(y)
    =-\frac{x^2}{2}-\frac{5 x}{8}-\frac{3}{4}+\frac{9}{8 x}+\frac{5}{4 x^2}-\frac{13}{8x^3}\;.\label{eqn:l=2phi2}
\end{align}
We emphasize here again that one should not directly read off the TLN as a coefficient of the $x^{-2}$ term in the above solution since there may be contamination coming from the subleading terms of the growing mode [see Eq.~(\ref{eq:psi_k_pert})].
Instead, one should compute the integral~\eqref{int_TLN} as
\begin{align}
    I[{\cal S}^{(2)}]
    =-\frac{1}{5}\int_1^x \de y\,\tilde{\psi}_{\rm RW}^+(y)\mathcal{S}^{(2)}(y)
    =-\frac{x^4}{10}-\frac{3x^3}{20}-\frac{x^2}{5}+\frac{27x}{20}+\frac{7}{20}-\frac{16}{5x}+\frac{39}{20x^2}\;,
\end{align}
whose constant part corresponds to the TLN:
    \begin{align}
    K_{\ell=2}^{(2)}=\frac{7}{20}\;.
    \end{align}
\end{itemize}

Likewise, it is straightforward to compute the TLN up to an arbitrary order in $\eta$.
For instance, the solution to Eq.~\eqref{eq:kth-order-eq} of ${\cal O}(\eta^3)$ and ${\cal O}(\eta^4)$ can be obtained as
    \begin{align}
    \tilde{\psi}_{\ell=2}^{(3)}(x)&=\frac{x^2}{2}+\frac{3 x}{4}+\frac{17}{16}-\frac{9}{16 x}-\frac{1}{x^2}-\frac{25}{16 x^3}-\frac{9}{4 x^4}-\frac{33}{16 x^5}+\frac{19}{16 x^6}\;, \label{eqn:l=2phi3} \\
    \tilde{\psi}_{\ell=2}^{(4)}(x)&=-\frac{x^2}{2}-\frac{7 x}{8}-\frac{23}{16}-\frac{29}{128 x}+\frac{7}{32 x^2}+\frac{71}{64 x^3}+\frac{41}{16 x^4}+\frac{409}{128 x^5}+\frac{7}{2 x^6}\nonumber\\
    &\quad +\frac{215}{64 x^7}+\frac{85}{32 x^8}-\frac{125}{128 x^9}\;, \label{eqn:l=2phi4}
    \end{align}
respectively.
By use of $\tilde{\psi}^{(i)}$ ($i=0,\cdots,k-1$) together with $V^{(j)}$ ($j=1,\cdots,k$) whose first few expressions are given in Eq.~\eqref{V_Hayward_eta_exp}, 
one can straightforwardly compute ${\cal S}^{(k)}$ with Eq.~\eqref{source_def}.
Then, the integral~\eqref{int_TLN} allows us to extract the contribution of ${\cal O}(\eta^k)$ to the TLN.
The result up to ${\cal O}(\eta^4)$ is therefore given by
\begin{align}
    K_{\ell=2}
    =\frac{7}{20}\eta^2-\frac{11}{20}\eta^3+\frac{2}{5}\eta^4+\cdots\;.\label{TLN-l=2}
\end{align}

\subsubsection{\texorpdfstring{$\ell=3$ case}{l=3 case}}
In the case of $\ell=3$, we have
\begin{align}
   \tilde{\psi}_{\rm RW}^+(x)&=x^4-\frac{5}{6}x^3=\tilde{\psi}_{\ell=3}^{(0)}\;,\\
   \tilde{\psi}_{\rm RW}^-(x)&=-252\left[ \left(x^4-\frac{5}{6}x^3\right)\log \left(1-\frac{1}{x}\right)+x^3-\frac{x^2}{3}-\frac{x}{12}-\frac{1}{36}-\frac{1}{120x}\right].
\end{align}
The analysis proceeds in the same way as in the $\ell=2$ case, so we just present results of the computation.
The solutions to Eq.~\eqref{eq:kth-order-eq} up to ${\cal O}(\eta^3)$ can be obtained as
    \begin{align}
    \tilde{\psi}_{\ell=3}^{(1)}(x)
    &=\frac{4 x^3}{3}+\frac{x^2}{12}+\frac{19 x}{12}-\frac{25}{12}+\frac{1}{4 x}\;, \\
    \tilde{\psi}_{\ell=3}^{(2)}(x)
    &=-\frac{4x^3}{3}+\frac{5x^2}{24}+\frac{3x}{16}+\frac{37}{12}-\frac{25}{48x}-\frac{11}{8x^2}+\frac{17}{16x^3}-\frac{1}{8x^4}\;, \\
    \tilde{\psi}_{\ell=3}^{(3)}(x)
    &=\frac{4x^3}{3}-\frac{x^2}{2}-\frac{x}{2}-\frac{109}{32}+\frac{121}{96x}+\frac{1}{8x^2}-\frac{569}{336x^3}+\frac{23}{168x^4}+\frac{43}{32x^5}-\frac{83}{96x^6}+\frac{3}{32x^7}\;.
    \end{align}
Thus, following the same procedure as for $\ell=2$, the TLN for $\ell=3$ up to ${\cal O}(\eta^4)$ is obtained as
    \begin{align}
        K_{\ell=3}
        =\frac{5}{42}\eta+\frac{1417}{504}\eta^2-\frac{1285}{1008}\eta^3+\frac{3713}{4032}\eta^4+\cdots\;.\label{TLN-l=3}
    \end{align}

\subsubsection{\texorpdfstring{$\ell=4$ case}{l=4 case}}

In the case of $\ell=4$, we have
\begin{align}
    \tilde{\psi}_{\rm RW}^+(x)&=x^5-\frac{3 x^4}{2}+\frac{15 x^3}{28}=\tilde{\psi}_{\ell=4}^{(0)}(x)\;,\\
    \tilde{\psi}_{\rm RW}^-(x)&=-7056 \left[\left(x^5-\frac{3 x^4}{2}+\frac{15 x^3}{28}\right)\log \left(1-\frac{1}{x}\right)+x^4-x^3+\frac{5}{42}x^2+\frac{x}{56}+\frac{1}{280}+\frac{1}{1680x}\right].
\end{align}
As opposed to the cases of $\ell=2$ and $\ell=3$, there is a logarithmic running of TLNs for $\ell=4$ as we will see below.

\begin{itemize}
\item {\it First order in $\eta$}.
The source term of ${\cal O}(\eta)$ is given by
    \begin{align}
        \mathcal{S}^{(1)}(x)&=-16 x^2+\frac{17 x}{2}-\frac{513}{28}+\frac{2895}{56 x}-\frac{234}{7 x^2}+\frac{45}{8 x^3}\;,
    \end{align}
from which we obtain the first-order solution as
    \begin{align}
        \tilde{\psi}_{\ell=4}^{(1)}(x)&=\int_1^\infty \de y\,G(x,y)\mathcal{S}^{(1)}(y)
        =2 x^4-\frac{37 x^3}{28}+\frac{57 x^2}{56}-\frac{681 x}{280}+\frac{237}{200}-\frac{23}{200 x}\;.\label{eq:1st-order-l=4}
    \end{align}
From the constant part of the integral~\eqref{int_TLN}, we have
\begin{align}
    K_{\ell=4}^{(1)}=\frac{23}{840}\;.
\end{align}

\item {\it Second order in $\eta$}.
The source term of ${\cal O}(\eta^2)$ is given by
\begin{align}
    \mathcal{S}^{(2)}(x)&=16 x^2-\frac{99 x}{4}-\frac{135}{56}-\frac{979}{16 x}+\frac{19881}{280 x^2}+\frac{6717}{2800 x^3}-\frac{2323}{200 x^4}-\frac{14683}{2800 x^5}+\frac{1959}{280 x^6}-\frac{483}{400 x^7}\;.
\end{align}
Then, the integral~\eqref{int_TLN} can be computed as
    \begin{align}
    I[{\cal S}^{(2)}]
    &=-\frac{2x^8}{9}+\frac{65x^7}{84}-\frac{101x^6}{126}+\frac{7933x^5}{5040}-\frac{105509x^4}{23520}+\frac{511039x^3}{100800}-\frac{89471x^2}{70560} \nonumber \\
    &\quad -\frac{211111x}{141120}-\frac{24}{25}\log x+\frac{110051}{50400}-\frac{379231}{235200x}+\frac{12107}{39200x^2}-\frac{23}{960x^3}\;,
    \end{align}
which involves a logarithmic term as well as a constant term.
Therefore, we have
    \begin{align}
    K_{\ell=4}^{(2),\log}=-\frac{24}{25}\;, \qquad
    K_{\ell=4}^{(2)}=\frac{110051}{50400}\;.
    \end{align}
\end{itemize}

Thus, we have found that the TLN for $\ell=4$ is given by
    \begin{align}
    K_{\ell=4}
    =\frac{23}{840}\eta+\left(\frac{110051}{50400}-\frac{24}{25}\log x\right)\eta^2+\cdots\;.\label{TLN-l=4}
    \end{align}
Such a logarithmic running of TLNs would appear at ${\cal O}(\eta^2)$ for $\ell\ge 4$ in general.

\section{Summary and Discussion}\label{sec:discussion}

In this paper, we have investigated the tidal deformability of black holes (BHs) within the effective field theory (EFT) with a timelike scalar profile~\cite{Mukohyama:2022enj,Mukohyama:2022skk}.
The tidal deformability of an object under static perturbations is often quantified by tidal Love numbers (TLNs), which affect the waveform of gravitational waves during the inspiral phase of a compact binary merger.
In general relativity (GR), a Kerr BH (i.e., the unique stationary and rotating vacuum BH solution in GR) is known to have vanishing TLNs~\cite{Damour:2009vw,Binnington:2009bb,Kol:2011vg,Hui:2020xxx,LeTiec:2020bos,LeTiec:2020spy,Chia:2020yla,Charalambous:2021mea}.
However, in gravitational theories beyond GR, the TLNs can be non-vanishing in general, which offers an interesting possibility for testing gravity with gravitational wave observations.
It should be noted that our EFT encompasses any scalar-tensor theories in principle, and hence it serves as an ideal framework to study the universal feature of TLNs in scalar-tensor theories.

We have focused on the TLNs associated with odd-parity perturbations about a static and spherically symmetric BH with a timelike scalar profile in the EFT, for which the master equation (i.e., the generalized Regge-Wheeler equation) was derived in \cite{Mukohyama:2022skk}.
We have first discussed the case of stealth Schwarzschild solution, where the generalized Regge-Wheeler equation for this background is equivalent to that in GR upon an appropriate rescaling, and hence the TLNs are vanishing as in the GR case.
For this reason, in Section~\ref{ssec:TLN_extraction}, we have focused mostly on the TLNs for non-stealth BHs.
Even in such a case, the generalized Regge-Wheeler equation for static (or zero-frequency) perturbations can be rewritten in the form where all the deviations from GR is absorbed into a modification of the effective potential [see Eq.~\eqref{eq:GRW-f-static}].
Our strategy is to solve this differential equation perturbatively with respect to $\eta$ (a parameter that characterizes the deviation from the Schwarzschild background) and then compute the TLNs from the perturbative solutions.
In doing so, as detailed in Section~\ref{ssec:TLN_extraction}, we have performed an analytic continuation of the multipole index~$\ell$ to generic complex values to uniquely determine the TLNs for integer values of $\ell$, and thus circumvented the mixing between the growing/decaying at spatial infinity.
As a demonstration of our general methodology, in Section~\ref{ssec:Hayward}, we have considered the Hayward solution as an example of non-stealth BHs.
We have explicitly computed the TLNs perturbatively in $\eta$ and clarified that the TLNs are non-vanishing in general.
Also, we have found that a logarithmic running of TLN would show up for $\ell\ge 4$.
It would be intriguing to study how the effect of non-vanishing TLNs (or the logarithmic running) shows up in the gravitational waveform, which we leave for future work.

There are several other possible future studies.
It would be important to confirm that our computation of TLNs based on the analytic continuation of $\ell$ is consistent with other approaches, e.g., worldline EFT~\cite{Ivanov:2022hlo,Ivanov:2022qqt}.
Moreover, in order to verify that we properly separate the external tidal field and the response to it, one has to perform a matching with post-Newtonian metric~\cite{Blanchet:2013haa,Gralla:2017djj}.
Another is to get more insight into the condition under which TLNs are vanishing by means of our EFT.
As mentioned above, we have shown that the TLNs are zero for the stealth Schwarzschild solution, but there can be other solutions where TLNs vanish in principle.
If this is the case, it would be nice to understand the physics and/or mathematics behind it.
It is also interesting to study the dissipation number~\cite{LeTiec:2020spy,LeTiec:2020bos,Chia:2020yla,Charalambous:2021mea,Ivanov:2022hlo}, which is another quantity that characterizes a frequency-dependent tidal response.
We hope to come back to these issues in the future.

\section*{Acknowledgements}
The authors would like to thank Takuya Katagiri, Tact Ikeda, and Vitor Cardoso for useful discussions.
This work was supported in part by World Premier International Research Center Initiative (WPI), MEXT, Japan.
The work of H.K.~was supported by JST (Japan Science and Technology Agency) SPRING, Grant No.\ JPMJSP2110.
The work of S.M.~was supported in part by JSPS (Japan Society for the Promotion of Science) KAKENHI Grant No.\ JP24K07017. 
The work of N.O.~was supported by JSPS KAKENHI Grant No.\ JP23K13111 and by the Hakubi project at Kyoto University.
The work of K.T.~was supported by JSPS KAKENHI Grant No.\ JP23K13101.
The work of V.Y.~was supported by JSPS KAKENHI Grant No.\ JP22K20367.

\appendix

\section{Construction of Green's function}\label{app:detail-Green}

In order to solve the differential equation~\eqref{eq:kth-order-eq}, it is convenient to use the Green's function that satisfies
\begin{align}
    \left[\hat{\mathcal{L}}^2 - f(x)V^{(0)}(x)\right]G(x,y) = f(x)\delta(x-y)\;, \label{eq:def-Green-func}
\end{align}
where we have defined the differential operator $\hat{\cal L} \equiv f(x)(\de/\de x)$ with $f(x)=1-1/x$.
Here, we present the construction of Green's function under an appropriate set of boundary conditions.
Let us choose $\tilde{\psi}_{\rm RW}^{\text{hor-reg}}(x)$ and $\tilde{\psi}_{\rm RW}^-(x)$ (see Section~\ref{ssec:pert_sol} for the definition) as two linearly independent homogeneous solutions to the above equation and impose the following ansatz:
\begin{align}\label{eq:green_ansatz}
    G(x,y) = C_1(x,y)\tilde{\psi}_{\rm RW}^{\text{hor-reg}}(x) + C_2(x,y)\tilde{\psi}_{\rm RW}^-(x)\;,
\end{align}
with the functions~$C_1$ and $C_2$ satisfying
    \begin{align}
    (\hat{\mathcal{L}}C_1)\tilde{\psi}_{\rm RW}^{\text{hor-reg}}(x) + (\hat{\mathcal{L}}C_2)\tilde{\psi}_{\rm RW}^-(x) = 0\;.
    \end{align}
Notice that the above condition on $C_1$ and $C_2$ guarantees that, when $\hat{\cal L}$ acting on $G(x,y)$, one only gets $C_1 \hat{\cal L} \tilde{\psi}_{\rm RW}^{\text{hor-reg}}(x) + C_2 \hat{\cal L} \tilde{\psi}_{\rm RW}^-(x)$.
Plugging the ansatz~(\ref{eq:green_ansatz}) into Eq.~\eqref{eq:def-Green-func} yields
    \begin{align}
    (\hat{\mathcal{L}}C_1)\hat{\mathcal{L}}\tilde{\psi}_{\rm RW}^{\text{hor-reg}}(x)+(\hat{\mathcal{L}}C_2)\hat{\mathcal{L}}\tilde{\psi}_{\rm RW}^-(x)=f(x)\delta(x-y)\;,
    \end{align}
and therefore we have
\begin{align}
    \left(\begin{array}{cc}
        \tilde{\psi}_{\rm RW}^{\text{hor-reg}}(x) & \tilde{\psi}_{\rm RW}^-(x) \\
        \hat{\mathcal{L}}\tilde{\psi}_{\rm RW}^{\text{hor-reg}}(x) & \hat{\mathcal{L}}\tilde{\psi}_{\rm RW}^-(x)
    \end{array}\right)\left(\begin{array}{c}
        \hat{\mathcal{L}}C_1  \\
        \hat{\mathcal{L}}C_2
    \end{array}\right) &= \left(\begin{array}{c}
        0  \\
        f(x)\delta (x-y) 
    \end{array}\right).
\end{align}
Note that the determinant of the $2\times2$ matrix in the left-hand side is nothing but the Wronskian of the two homogeneous solutions [see also Eq.~\eqref{Wronskian}]:
\begin{align}
    W\!\left[\tilde{\psi}_{\rm RW}^{\text{hor-reg}}(x),\tilde{\psi}_{\rm RW}^-(x)\right] \equiv \tilde{\psi}_{\rm RW}^{\text{hor-reg}}(x)\hat{\mathcal{L}} \tilde{\psi}_{\rm RW}^-(x) - \tilde{\psi}_{\rm RW}^-(x) \hat{\mathcal{L}} \tilde{\psi}_{\rm RW}^{\text{hor-reg}}(x)=-(2\ell+1)\;.
\end{align}
As a result, we obtain
\begin{align}
    \frac{\de}{\de x}\left(\!\begin{array}{c}
        C_1  \\
        C_2
    \end{array}\!\right) &= -\frac{1}{2\ell+1}\left(\begin{array}{c}
        -\tilde{\psi}_{\rm RW}^-(x)\delta(x-y)  \\
        \tilde{\psi}_{\rm RW}^{\text{hor-reg}}(x)\delta(x-y) 
    \end{array}\right),
\end{align}
and hence
\begin{equation}
    \begin{split}
        C_1(x) &= \frac{1}{2\ell+1}\int_\infty^x \de x'\,\tilde{\psi}_{\rm RW}^-(x')\delta(x'-y)
        =-\frac{1}{2\ell+1}\tilde{\psi}_{\rm RW}^-(y)\Theta(y-x)\;,\\
        C_2(x) &= -\frac{1}{2\ell+1}\int_1^x \de x'\,\tilde{\psi}_{\rm RW}^{\text{hor-reg}}(x')\delta(x'-y)
        =-\frac{1}{2\ell+1}\tilde{\psi}_{\rm RW}^{\text{hor-reg}}(y)\Theta(x-y)\;.
    \end{split}
\end{equation}
The lower bounds of the integrals above have been fixed from the boundary conditions that the field is regular (i.e., $\tilde{\psi}_{\rm RW}^-$ is absent) at $x=1$ and $\tilde{\psi}^{(k)}(x) \sim {\cal O}(x^{\ell})$ (i.e., $\tilde{\psi}_{\rm RW}^{\text{hor-reg}}$ is absent) at spatial infinity [see the comment below Eq.~\eqref{GreenF_def}].

Thus, the Green's function is given by
    \begin{align}
    G(x,y)=-\frac{1}{2\ell+1}\left[\tilde{\psi}_{\rm RW}^-(x)\tilde{\psi}_{\rm RW}^{\text{hor-reg}}(y)\Theta(x-y)+\tilde{\psi}_{\rm RW}^{\text{hor-reg}}(x)\tilde{\psi}_{\rm RW}^-(y)\Theta(y-x)\right].
    \end{align}
Note that the Green's function is symmetric about $x$ and $y$.

\section{Parameterized TLN formalism}
\label{app:pTLN}

In the main text, we have started from our EFT and studied TLNs of odd-parity perturbations based on the generalized RW equation~\eqref{eq:GRW-f-static}, where the modification from GR is encoded in the effective potential~$V(x)$.
A more phenomenological approach is to consider the following parameterized modification of $V(x)$, without specifying the underlying theory or the background spacetime~\cite{Katagiri:2023umb}:
    \begin{align}
    V(x)=V^{(0)}(x)+\eta\sum_{j=3}^{N_1}\frac{v_j}{x^j}\;, \qquad
    V^{(0)}(x)=\frac{\ell(\ell+1)}{x^2}-\frac{3}{x^3}\;,
    \label{V_parameterized}
    \end{align}
where $V^{(0)}$ represents the standard RW potential and $\eta$ is a small constant.
Here, we have assumed that the modification from GR starts at $x^{-3}$, so that the leading asymptotic behavior of the perturbations is not affected by the modification characterized with the constants~$v_j$.
Then, up to the first order in $\eta$, the TLNs can be expressed in the form~\cite{Katagiri:2023umb}
    \begin{align}
    K_\ell=\eta\sum_{j=3}^{N_1}v_j e_j+{\cal O}(\eta^2)\;,
    \label{pTLN}
    \end{align}
where the ($\ell$-dependent) constants~$e_j$ play the role of a basis for the TLNs.
When the logarithmic running of the TLNs shows up, we write
    \begin{align}
    K_\ell^{\log}=\eta\sum_{j=3}^{N_1}v_j e_j^{\log}+{\cal O}(\eta^2)\;,
    \end{align}
with the constants~$e_j^{\log}$ being a basis for the logarithmic running.
Once the bases are obtained, one can compute the TLN (and its logarithmic running) at ${\cal O}(\eta)$ from any given effective potential of the form~\eqref{V_parameterized} via Eq.~\eqref{pTLN}.

Let us now find the expression of $e_j$ and $e_j^{\log}$ based on our methodology developed in Section~\ref{ssec:extraction_TLN}.
The TLN (and its logarithmic running) at ${\cal O}(\eta)$ can be extracted from the constant part of the integral~\eqref{int_TLN}, which in the present case reads
    \begin{align}
    I[{\cal S}^{(1)}]=-\frac{1}{2\ell+1}\sum_{j=3}^{N_1}v_j\int_1^x {\rm d}y\,y^{-j}\left[\tilde{\psi}_{\rm RW}^+(y)\right]^2\;.
    \end{align}
By use of the expression for $\tilde{\psi}_{\rm RW}^+$ given in Eq.~\eqref{psi+}, we obtain
    \begin{align}
    e_j&=\frac{1}{2\ell+1}\sum_{(m,p)\in\mathbb{A}^2\backslash\mathbb{B}}\frac{a_m^+ a_p^+}{2\ell-j-m-p+3}\;, \qquad
    e_j^{\log}=-\frac{1}{2\ell+1}\sum_{(m,p)\in\mathbb{A}^2\cap\mathbb{B}}a_m^+ a_p^+\;, \label{ej^log}
    \end{align}
where we have defined
    \begin{align}
    \mathbb{A}\equiv \left\{m\in\mathbb{Z}\mathrel{}\middle|\mathrel{}0\le m\le \ell-2\right\}, \qquad
    \mathbb{B}\equiv \left\{(m,p)\in\mathbb{Z}^2\mathrel{}\middle|\mathrel{}2\ell-j-m-p+3=0\right\}.
    \end{align}
Here, $\mathbb{B}\ne\emptyset$ if and only if $7\le j\le 2\ell+3$, and hence $e_j^{\log}\ne 0$ only for these values of $j$.
One can obtain the following generating function of $e_j^{\log}$:
    \begin{align}
    \sum_{j\ge 3}e_j^{\log}z^{j-1}
    =-\frac{1}{2\ell+1}\left[\tilde{\psi}_{\rm RW}^+(z)\right]^2.
    \end{align}
Regarding $e_j$, we have
    \begin{align}
    e_j=\sum_{n\ge 3\,\&\,n\ne j}\frac{e_n^{\log}}{j-n}\;,
    \end{align}
which holds for any $j\ge 3$, including $7\le j\le 2\ell+3$.
The values of $e_j$ and $e_j^{\log}$ up to $\ell=4$ are summarized in the following \hyperref[pTLN_basis]{Table}.
    \begin{table}[H]
    \renewcommand\thetable{\!\!}
    \centering
    \caption{Basis for the (logarithmic running of) TLNs up to $\ell=4$.}
    \label{pTLN_basis}
    \vskip2mm
    \begin{tabular}{|c||c|c|}
    \hline
    $\ell=2$ & $j\ge 3$ \& $j\ne 7$ & $j=7$ \\
    \hline\hline
    $e_j$ & $\displaystyle -\frac{1}{5(j-7)}$ & $0$ \rule[-5mm]{0mm}{12mm} \\
    \hline
    $e_j^{\log}$ & $0$ & $\displaystyle -\frac{1}{5}$ \rule[-5mm]{0mm}{12mm} \\
    \hline
    \end{tabular}
    \hskip2mm
    \begin{tabular}{|c||c|c|c|c|}
    \hline
    $\ell=3$ & $j\ge 3$ \& $j\ne 7,8,9$ & $j=7$ & $j=8$ & $j=9$ \\
    \hline\hline
    $e_j$ & $\displaystyle -\frac{36-5j+j^2}{252(j-7)(j-8)(j-9)}$ & $\displaystyle -\frac{1}{6}$ & $\displaystyle \frac{11}{252}$ & $\displaystyle \frac{95}{504}$ \rule[-5mm]{0mm}{12mm} \\
    \hline
    $e_j^{\log}$ & $0$ & $\displaystyle -\frac{25}{252}$ & $\displaystyle \frac{5}{21}$ & $\displaystyle -\frac{1}{7}$\rule[-5mm]{0mm}{12mm} \\
    \hline
    \end{tabular}\\
    \vskip2mm
    \begin{tabular}{|c||c|c|c|c|c|c|}
    \hline
    $\ell=4$ & $j\ge 3$ \& $j\ne 7,8,9,10,11$ & $j=7$ & $j=8$ & $j=9$ & $j=10$ & $j=11$ \\
    \hline\hline
    $e_j$ & $\displaystyle -\frac{2712-890j+203j^2-10j^3+j^4}{7056(j-7)(j-8)(j-9)(j-10)(j-11)}$ & $\displaystyle -\frac{13}{168}$ & $\displaystyle \frac{4393}{21168}$ & $\displaystyle -\frac{1625}{14112}$ & $\displaystyle -\frac{1265}{7056}$ & $\displaystyle \frac{1885}{9408}$ \rule[-5mm]{0mm}{12mm} \\
    \hline
    $e_j^{\log}$ & $0$ & $\displaystyle -\frac{25}{784}$ & $\displaystyle \frac{5}{28}$ & $\displaystyle -\frac{31}{84}$ & $\displaystyle \frac{1}{3}$ & $\displaystyle -\frac{1}{9}$ \rule[-5mm]{0mm}{12mm} \\
    \hline
    \end{tabular}
    \end{table}
Let us comment on the comparison between our results and those obtained in \cite{Katagiri:2023umb}.
As we have explained in detail in Section~\ref{sec:analysis}, for $\ell\in\mathbb{Z}_{\ge 2}$ corresponding to physical multipoles, there is an ambiguity in the definition of TLNs as the separation between the growing/decaying modes at spatial infinity is not unique.
In our case, we have circumvented this problem by performing an analytic continuation of $\ell$ to generic complex values.
On the other hand, in \cite{Katagiri:2023umb}, the authors only studied the case where $\ell$ is an integer and read off the TLN as the coefficient of the $x^{-\ell}$ term in the large-$x$ expansion of the field, which implies that the ambiguity may remain in their results.
For $j\ge 2\ell+4$, our analysis shows that there is no contamination in the $x^{-\ell}$ term from the growing mode as the series expansion of the growing mode does not contain such a term.\footnote{Indeed, for $j\ge 2\ell+4$, one can show that the first line of Eq.~\eqref{eq:inhomo-generic}, which is the only contribution to the growing mode, behaves as $x^{-\ell-1}[1+{\cal O}(x^{-1})]$ and hence does not affect the $x^{-\ell}$ term. (Note that we now assume $\ell\in\mathbb{Z}_{\ge 2}$.)}
Therefore, as it should, we have reproduced their results of $e_j$ for these values of $j$.
Also, our results of $e_j^{\log}$ are consistent with theirs.
However, the contamination indeed happens for smaller values of $j$.
They showed that the $x^{-\ell}$ term is absent in the series expansion of the field for $3\le j\le 6$, but we have found that this is a result of a cancellation between the contributions from the growing/decaying modes.

We now apply the above formalism to the case of the Hayward background in our EFT, for which the non-vanishing expansion coefficients of $V(x)$ are given by [see Eq.~\eqref{V_Hayward_eta_exp}]
    \begin{align}
    v_3=-\ell(\ell+1)+4\;, \qquad
    v_4=-\ell(\ell+1)+\frac{9}{2}\;, \qquad
    v_5=-\ell(\ell+1)-13\;, \qquad
    v_6=\frac{21}{2}\;.
    \end{align}
It is straightforward to confirm that Eq.~\eqref{pTLN} with the basis for TLNs presented in the above \hyperref[pTLN_basis]{Table} reproduces the results in Section~\ref{ssec:Hayward} at ${\cal O}(\eta)$.
It should be noted that we find $K_{\ell=2}={\cal O}(\eta^2)$ as a result of a non-trivial cancellation among contributions of different values of $j$.
Note also that the logarithmic running is absent up to ${\cal O}(\eta)$ for this particular background.

\section{\texorpdfstring{Exact solution of $\ell=2$ mode on the Hayward background}{Exact solution of l=2 mode on the Hayward background}}\label{app:exact-solutions}

In this Appendix, we present an exact solution for the $\ell=2$ static perturbation on the Hayward background.
We introduce $D_T(x) \equiv (1+\alpha_T)^{1/4}/x$, so that the effective potential~$\tilde{V}(x)$ in the generalized RW equation~\eqref{eq:GRW-x} takes the following simple form:
    \begin{align}
        \tilde{V}(x)=&[\ell(\ell+1)-2]\sqrt{\frac{A}{B}}\,D_T(x)^2+\frac{1}{D_T(x)}\frac{\de}{\de x}\left( F\cdot \frac{\de}{\de x}D_T(x)\right).\label{eq: GRW-potential-dimless-2}
\end{align}
Then, in terms of a new variable
\begin{align}
    R(x) \equiv \frac{\tilde{\Psi}(x)}{D_T(x)}\;,
    \label{R_def}
\end{align}
Eq.~\eqref{eq:GRW-x} can be rewritten as
\begin{equation}
    \frac{\de}{\de x}\left(F(x)D_T^2(x) \frac{\de}{\de x}R(x)\right)-[\ell(\ell+1)-2]\sqrt{\frac{A}{B}}\,D_T^4(x)R(x)=0\;.\label{eqn:GRW-another}
\end{equation}
Note that we have not assumed any particular background so far.
In the case of Hayward background, we have
\begin{align}
    D_T(x)
    =\sqrt{\frac{x}{x^3+\eta}}\;.
\end{align}

One can show that the two linearly independent solutions to Eq.~\eqref{eqn:GRW-another} for $\ell = 2$ are given by
\begin{equation}\label{exact-l=2}
\begin{split}
    R_1(x)&\propto {x^4+4\eta x}\;,\\
    R_2(x)&\propto R_1(x)\int_\infty^x \de x'\frac{(x'^3+\eta)^2}{(x'-1)[x'^3+\eta(x'+1)(x'^2+1)]R_1^2(x')}\;.
\end{split}
\end{equation}
The above exact solutions can be translated into the language of $\tilde{\psi}$ used in the main text.
By using the relation~$\tilde{\psi}(x)=\sqrt{F(x)/f(x)}\,\tilde{\Psi}(x)$ with Eq.~\eqref{R_def}, the two independent solutions of $\tilde{\psi}$ corresponding to $R_1(x)$ and $R_2(x)$ are obtained as
    \begin{align}
    \tilde{\psi}_1&=\sqrt{\frac{x[x^3+\eta(x+1)(x^2+1)]}{1+\eta}}\frac{x^4+4\eta x}{x^3+\eta}\;, \\
    \tilde{\psi}_2&=-5\tilde{\psi}_1\int_\infty^x \de x'\frac{x'}{(x'-1)\tilde{\psi}_1^2(x')}\;,
    \end{align}
respectively, where the coefficients have been chosen so that the coefficients of the leading order in the $1/x$-expansion are unity, i.e.,
    \begin{align}
    \tilde{\psi}_1&=x^3+\frac{\eta x^2}{2(1+\eta)}+\frac{\eta(4+3\eta)x}{8(1+\eta)^2}+{\cal O}(x^0)\;, \\
    \tilde{\psi}_2&=\frac{1}{x^2}+\frac{5+3\eta}{6(1+\eta)x^3}+\frac{120+154\eta+63\eta^2}{168(1+\eta)^2x^4}+{\cal O}(x^{-5})\;.
    \end{align}
Indeed, in the limit~$\eta\to 0$, $\tilde{\psi}_1$ and $\tilde{\psi}_2$ reduce to $\tilde{\psi}_{\rm RW}^+$ and $\tilde{\psi}_{\rm RW}^-$ in Eqs.~\eqref{psi+_l=2} and \eqref{psi-_l=2}, respectively.
Also, $\tilde{\psi}_1$ corresponds to the solution that is regular at the horizon, and its perturbative expansion in $\eta$ reproduces Eqs.~\eqref{l=2-1st-order}, \eqref{eqn:l=2phi2}, \eqref{eqn:l=2phi3}, and \eqref{eqn:l=2phi4}.
It is important to note that if, on the other hand, one is able to obtain an exact solution to Eq.~(\ref{eqn:GRW-another}) for generic $\ell$, one would then analytically continue $\ell$ to complex numbers and uniquely extract the TLNs. Since it is complicated to do so, we will just follow the procedure described in Section~\ref{ssec:extraction_TLN}.

In order to apply our methodology to obtain the TLNs from the exact solution~$\tilde{\psi}_1$, we recast Eq.~\eqref{eq:GRW-f-static} into the form,
    \begin{align}
    \left[f(x)\frac{\de}{\de x}\left(f(x)\frac{\de}{\de x}\right)-f(x)V^{(0)}(x)\right]\tilde{\psi}_1(x)=f(x)\left[V(x)-V^{(0)}(x)\right]\tilde{\psi}_1(x)
    \equiv f(x){\cal S}_1(x)\;.
    \end{align}
Here, the differential operator acting on $\tilde{\psi}_1$ is of the standard Regge-Wheeler type, for which we have constructed the Green's function~(\ref{Green_func}) and argued that the TLNs can be extracted from the integral~\eqref{int_TLN}.
Note that one should replace the source term by ${\cal S}_1(y)$ defined above, with the expression of $V$ given in Eq.~\eqref{V-hayward-full}.
Namely, one should extract the constant part of the $1/x$-expansion of the following integral:
    \begin{align}
    I[{\cal S}_1]
    &=-\frac{1}{5}\int_1^x \de y\,\tilde{\psi}_{\rm RW}^+(y)\left[V(y)-V^{(0)}(y)\right]\tilde{\psi}_1(y) \nonumber \\
    &=\frac{\eta x^{5/2}(x-1)}{10(x^3+\eta)^2\sqrt{(1+\eta)[x^3+\eta(x+1)(x^2+1)]}} \nonumber \\
    &\quad \times\left[x^6(x^2+2x+21)+\eta x^3(18x^3+23x^2+28x+33)+4\eta^2(x^2+2x+3)\right].
    \label{int_TLN_exact}
    \end{align}
As a result, we obtain
    \begin{align}
    K_{\ell=2}&=\frac{\eta^2(448+1536\eta+1472\eta^2+323\eta^3-96\eta^4)}{1280(1+\eta)^5} \label{eq:Love_exact_l=2} \\
    &=\frac{7}{20}\eta^2-\frac{11}{20}\eta^3+\frac{2}{5}\eta^4+\frac{323}{1280}\eta^5-\frac{2031}{1280}\eta^6+\cdots\;,
    \end{align}
which precisely reproduces the perturbative result~\eqref{TLN-l=2}, as long as $|\eta| < 1$.
Interestingly, we note that the formula~(\ref{eq:Love_exact_l=2}) is exact and holds even if $\eta \ge 1$.
(Recall that we assume $\eta>-1/4$ so that the effective metric for the odd mode has a horizon.)
In fact, one can regard Eq.~(\ref{eq:Love_exact_l=2}) as a resummation of the perturbative result, computed in the main text.
Note also that the logarithmic running of TLNs for the modes with $\ell = 2$ is absent at any order in $\eta$.

\bibliographystyle{utphys}
\bibliography{bib}

\end{document}